\documentclass[floatfix,twocolumn,preprintnumbers,amsmath,amssymb,superscriptaddress]{revtex4-2}
\pdfoutput=1 
\usepackage{amsmath,amssymb,eucal,graphicx,float,epstopdf,xparse}

\usepackage[utf8]{inputenc}
\usepackage{comment}
\usepackage{subcaption}

\def \equi#1{\mathrel{\mathop{\kern 0pt\sim}\limits_{#1}}} 
\usepackage{xcolor}

\usepackage{subcaption}
\usepackage{dsfont}
\usepackage[pdfencoding=auto]{hyperref}
\usepackage{bookmark}

\newcommand{\dw}{d_\text{w}}
\newcommand{\df}{d_\text{f}}
\newcommand{\Sim}[1]{\mathrel{\mathop{\kern 0pt\sim}\limits_{#1}}}
\usepackage[shortlabels]{enumitem}

\begin{document}

\title{
Flips Reveal the Universal Impact of Memory on Random Explorations}

\author{J. Brémont}
\affiliation{ 
Laboratoire de Physique Théorique de la Matière Condensée, CNRS, UPMC, 4 Place Jussieu, 75005 Paris, France 
}
\affiliation{ 
Laboratoire Jean Perrin, CNRS, UPMC, 4 Place Jussieu, 75005 Paris, France 
}
\author{L. Régnier} \affiliation{ 
Laboratoire de Physique de l’École Normale Supérieure, CNRS, ENS \& PSL University, 75005 Paris, France
}
\affiliation{ 
Laboratoire Jean Perrin, CNRS, UPMC, 4 Place Jussieu, 75005 Paris, France 
}

\author{A. Barbier--Chebbah}
\affiliation{ 
Decision and Bayesian Computation, USR 3756 (C3BI/DBC) \& Neuroscience
department CNRS UMR 3751, Institut Pasteur, CNRS, Paris, France
}%
\affiliation{ 
    Epimethée, Inria, Paris, France.
}%

\author{R. Voituriez}\affiliation{ 
Laboratoire Jean Perrin, CNRS, UPMC, 4 Place Jussieu, 75005 Paris, France
}%

\author{O. Bénichou}\affiliation{ 
Laboratoire de Physique Théorique de la Matière Condensée, CNRS, UPMC, 4 Place Jussieu, 75005 Paris, France 
}

\date{\today}

\begin{abstract}
  Quantifying space exploration is a central question in random walk theory, with direct applications ranging from animal foraging \cite{ants,animals}, diffusion-limited reactions \cite{benichouGeometrycontrolledKinetics}, and intracellular transport \cite{Alessandro:2021,Stadler:2017} to stock markets \cite{bouchaudTheoryFinancial,drawdown,options}. In particular, the explored domain by one or many simple memoryless (or Markovian) random walks has received considerable attention \cite{rudnick,regnierUniversalExplorationDynamics2023,larraldeTerritoryCovered,hughes}. However, the physical systems mentioned above typically involve significant memory effects \cite{tarasovFractionalDynamics,Phillip:2021,mandelbrotFractionalBrownian,cellattract,lucas}, and so far, no general framework exists to analyze such systems. We introduce the concept of a \emph{flip}, defined most naturally in one dimension, where the visited territory is $[x_{\rm min}, x_{\rm max}]$: a flip occurs when, after discovering a new site at $x_{\rm max}$, the walker next discovers $x_{\rm min} - 1$ instead of $x_{\rm max} + 1$ (and vice-versa).
  While it reduces to the classical splitting probability in Markovian systems \cite{Bray:2013,hughes,majumdar}, we show that the flip probability serves as a key observable for quantifying the impact of memory effects on space exploration.
  Here, we demonstrate that the flip probability follows a strikingly simple and universal law: it decays inversely with the number of sites visited, as \(1/n\), independently of the underlying stochastic process. We confirm this behavior through simulations across paradigmatic non-Markovian models and observe it in real-world systems, without relying on model assumptions, including biological tracer motion \cite{Stadler:2017,Krapf:2019,Phillip:2021}, DNA sequences \cite{GenBank,gc1}, and financial market dynamics. Finally, we reveal the physical mechanism behind this universality and show how it extends to higher-dimensional and fractal domains. Our determination of universal flip statistics lay the groundwork for understanding how memory effects govern random explorations.
\end{abstract}

\maketitle
Fluctuations and reversals are fundamental features of complex systems. A stock price climbs to a record high—then crashes. A foraging animal searches one side of its habitat—then turns back in search of more food. A computer network handles a surge of data—then buckles under the load. At the heart of these diverse examples, commonly modeled as random walks (RWs) \cite{bouchaudTheoryFinancial,lelandSelfsimilarNatureEthernet1994,stochbookfin,boyerScalefreeForaging}, lies a simple, general question: as a system randomly explores new territory within its phase space, how often does it reverse the direction of its exploration?

\begin{figure}[h!]
  \centering
  \begin{subfigure}[t]{\columnwidth}
      \centering
      \includegraphics[width=\textwidth]{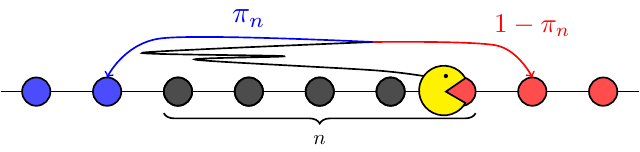}
      \caption{Flips in the exploration setting. Here, the walker (in yellow) consumes its $n^\text{th}$ unit of food, which is red: previously eaten food is in grey. A flip occurs when the next unit eaten is blue, which happens with probability $\pi_n$.}        
      \label{fig:flip-analogy}
  \end{subfigure}
  
  \vspace{1em} 

  \begin{subfigure}[t]{\columnwidth}
      \centering
      \includegraphics[width=.7\textwidth]{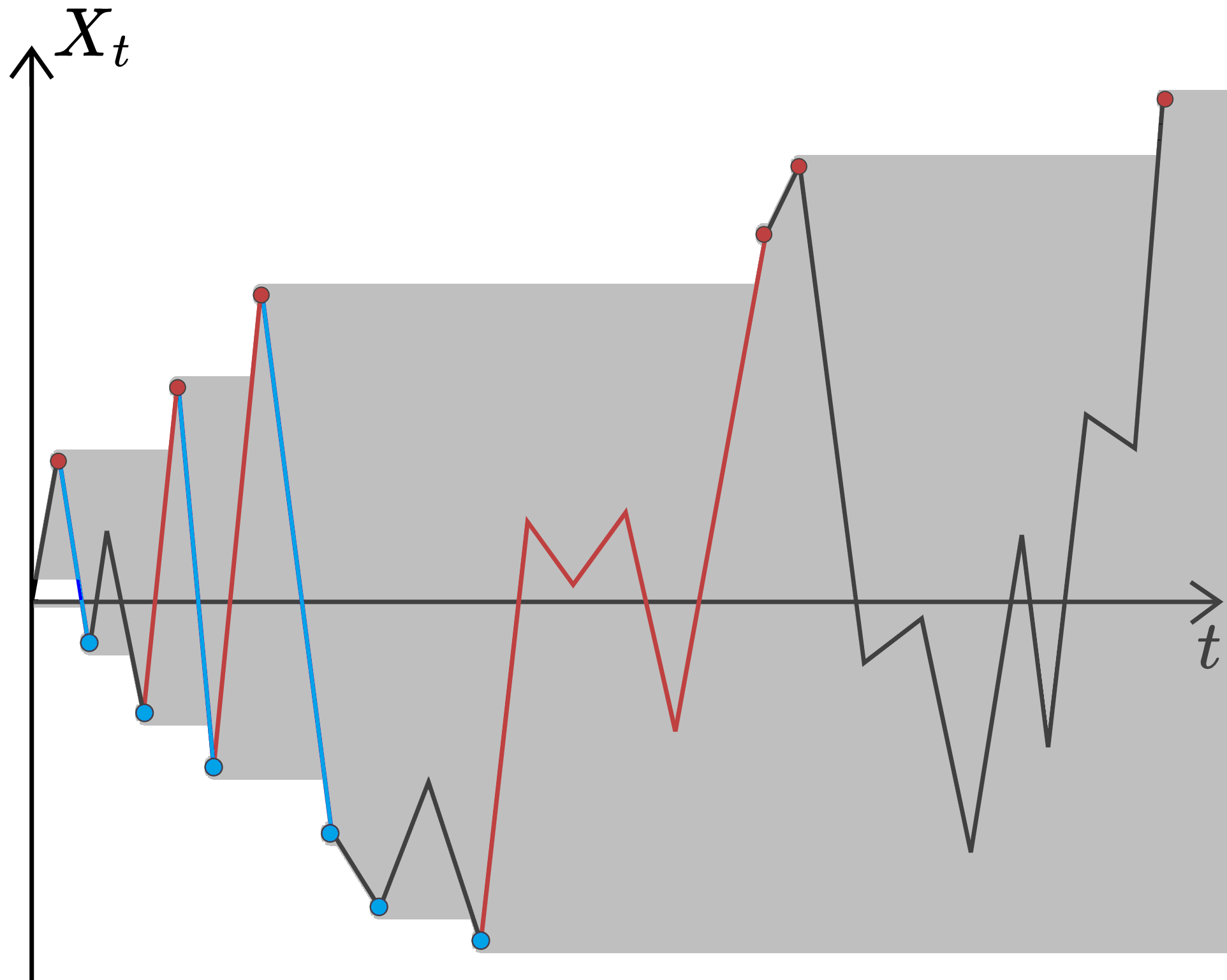}
      \caption{Flips in a time series $X_t$. Flips are identified as transitions between maxima and minima: downward flips (from maxima to minima) are marked in blue, and upward flips (from minima to maxima) are marked in red. The grey region shows the visited territory, i.e values already reached by $X_t$.
      }
      \label{fig:flip-timeseries}
  \end{subfigure}
  
  \caption{Illustration of the concept of a flip in two physically relevant settings. (a) Foraging analogy of flips; (b) flips in a generic time series.}
  \label{fig:Illus1}
\end{figure}

We introduce the concept of a \emph{flip} (see Fig.~\ref{fig:Illus1}) to capture this essential phenomenon. Defined most naturally in one dimension (see below for a generalization to higher dimensions), flips are transitions from a maximum to a minimum, or vice-versa, thus providing a direct measure of persistence in exploration. For Markovian processes, flip probabilities reduce to classical splitting probabilities—a well-understood quantity \cite{hughes,krapivskyKineticView}. But real-world phenomena are rarely memoryless \cite{tarasovFractionalDynamics,cellattract,saxtonAnomalousDiffusion,barkaiAreaBessel,Alessandro:2021,lucas,boyerRandomWalks}. From foraging patterns to stock market dynamics and intracellular transport, systems often carry the imprint of their past: their future depends not only on their current state but on their entire history.

This brings us to the central question of this article: how does memory shape the process of exploration ?  We address this fundamental problem by developing an analytical framework that allows us to compute the probability $\pi_n$ of a flip when the system has aged, so that the visited territory has size $n$. Our central result is a remarkably simple and universal scaling law: in one dimension, the probability of a flip scales as $\pi_n \propto 1/n$, as given in Eq. \eqref{pin} (see below for the extension to higher dimensions). 
This universality is particularly striking because it arises in memory-driven dynamics, where one typically expects complex, model-specific behavior \cite{Bray:2013,Toth:1995,baraviFirstPassage}. Indeed, the classical splitting probability, which quantifies the chance of crossing from one boundary of a fixed region to another, is known to depend sensitively on microscopic details of the process \cite{majumdar}. In stark contrast, flips occur when the walker traverses from one edge of its own dynamically explored region to the other. This moving, history-defined domain introduces memory and aging into the observable itself. Paradoxically, we find that this added complexity imposes simplicity: it erases process-specific dependencies and enforces a universal $1/n$ scaling. That this law emerges within the walker's self-generated geometry rather than in a fixed domain may explain why it has not been previously identified. \par 
The universality of our scaling law, which governs flips in effectively one-dimensional systems as varied as intracellular transport \cite{Phillip:2021,Krapf:2019,Krapf:2019}, nucleotide distribution along DNA strands \cite{GenBank}, and financial indices, is supported by both numerical simulations (Fig.~\ref{fig:splitting}) and experimental observations (Fig.~\ref{fig:pi_n_real}), underscoring its broad applicability across diverse natural systems. Finally, we demonstrate that our universal law extends beyond one-dimensional exploration, applying also to RWs in fractal media and higher-dimensional spaces. This further highlights the broad generality and far-reaching relevance of the flip-based framework.
\par 

To derive theoretically the general properties of flip probabilities,  we consider a general one-dimensional, non-Markovian, symmetric RW $(X_t)_{t=0,1,\ldots}$ in discrete time, with $X_0 = 0$. The dynamics are described by $X_{t+1} = X_t + \eta_t$, where the increments $(\eta_t)_t$ may exhibit correlations. We assume that $X_t$ converges to a scale-invariant process over long times, characterized by its walk dimension $\dw$ \cite{Bouchaud:1990}, meaning that the scaled random variable $X_t / t^{1/\dw}$ becomes asymptotically independent of $t$. 
In the following, we  are interested in the exploration process of the RW. To quantify exploration properties, we consider the integer lattice $\mathbb{Z}$ as a collection of sites, which are deemed visited by the RW when crossed \footnote{See \eqref{pin-compact} for the alternative arrival prescription, which involves holes in the visited region.}. The crossing prescription ensures that the visited sites form an interval.   
A relevant example of an exploration process is foraging, where sites are interpreted as units of food, represented by filled balls, eaten by the walker upon crossing (see Fig.~\ref{fig:Illus1}). To account for asymmetry in space exploration, we color food on positive (resp. negative) sites red (resp. blue). \par

We introduce and answer the following questions. When the walker has eaten $n$ units of food, we can assume, by symmetry, that the last unit consumed is red.  What is the probability $ \pi_n $ that the RW \textit{flips}, meaning the next unit consumed is blue (Fig.~\ref{fig:Illus1}) ? We propose $\pi_n$ as a novel metric for quantifying random explorations. As underlined in the introduction, for Markovian RWs, determining $ \pi_n $ is straightforward. Since the process is Markovian, the flip probability reduces to the classical splitting probability $q_n$: the probability that, starting from site $k$, the walker reaches site $-(n - k)$ before $k + 1$, independent of the past trajectory. This observable has led to multiple studies in the RW literature, and large-$n$ behavior is known explicitly and holds even for non-Markovian systems \footnote{We write $A_n \sim B_n$ if $A_n / B_n \to 1$ as $n \to \infty$.} : $q_n \propto n^{-\phi}$ \cite{majumdar}. Crucially, this decay is process-dependent: the exponent $\phi \equiv \dw \theta$, where $\theta$ is the persistence exponent, which characterizes the algebraic decay $S_x(t) \propto t^{-\theta}$ of the probability that the RW remains below a threshold $x > 0$ up to time $t$ \cite{Bray:2013}. For symmetric Markovian processes, $\theta = \frac{1}{2}$, as given by the Sparre-Andersen theorem, which in particular yields the well-known result $q_n \propto 1/n$ for such processes.
Our central result is that memory drastically impacts the statistics of flips in a remarkably simple way. Specifically, the probability $\pi_n$ follows a fully universal scaling law $\pi_n \sim A/n$, in striking contrast with the classical splitting probability $q_n$, for which the decay explicitly depends on $\phi$. The prefactor $A$, however, remains process-specific, serving as a fingerprint of the underlying microscopic dynamics.

We first determine, relying on a simple scaling analysis, the scaling with $n$ of the probability $\pi_n$ of a flip occurring after $n$ sites have been visited, that is, the probability of visiting the $(n+1)^{\text{th}}$ site on the other side of the $n^\text{th}$ visited site. See the discussion leading to Eq. \eqref{pin-compact} for a more refined argument which provides the physical mechanism responsible for the universal $1/n$ behavior. First, given the scale-invariant nature of the problem, we expect a power-law decay $\pi_n \propto n^{-\alpha}$, where $\alpha>0$. After $n$ sites have been visited, an average of $F_n \equiv \sum_{i=1}^n \pi_i$ flips have occurred. Since the RW is recurrent, we must have $F_n \to \infty$: otherwise, the RW would eventually stop visiting sites on one side. This implies that the exponent $\alpha\leq 1$. Let us assume that $\alpha<1$. Then $F_n \propto n^{1-\alpha}$ flips have occurred after $n$ sites have been visited, and, conversely, at the $k^\text{th}$ flip, $k^{1/(1-\alpha)}$ sites have been visited. 
Note that the only relevant length is the size of the visited domain. Hence, due to scale invariance, the number of visited sites between the $(k+1)^{\text{th}}$ and $k^{\text{th}}$ flips, of the order of $(k+1)^{1/(1-\alpha)} - k^{1/(1-\alpha)} \propto k^{\alpha/(1-\alpha)}$, must be proportional to the length $k^{1/(1-\alpha)}$ of the entire visited interval at the $k^\text{th}$ flip. This implies $\frac{\alpha}{1-\alpha}=\frac{1}{1-\alpha}$, which contradicts $\alpha<1$. This finally yields that $\alpha=1$, and thus the general behavior $\pi_n \sim A/n$ \footnote{more precisely, our analysis shows that $F_n\propto \ln n $, meaning the number of sites visited at the $k^\text{th}$ flip is exponential with $k$. This is consistent with the proportionality between the number of newly and previously visited sites.}. \par 
Remarkably, the universal \(1/n\) decay emerges independently of microscopic details, while the prefactor \(A\) is model-dependent. In contrast to the scaling argument yielding the $1/n$ decay, determining \(A\) requires technical, process-specific calculations, outlined in the Supplementary Information (SI). To provide a comprehensive framework, we classify memory in scale-invariant processes into three broad classes \cite{Regnier:2023}, covering the principal types of memory-driven dynamics that are most prominently studied and recognized as physically relevant in the literature. See the Methods section for the explicit description of these processes. First, we consider the class (I) of RWs for which flips are independent of each other (or at least asymptotically, at large number of visited sites $n$). Beyond Markov processes, we show in SI that this class includes important non-Markovian examples such as Lévy Walks \cite{zaburdaevLevyWalks,bacteria}, the Random Acceleration Process (RAP) \cite{meersonGeometricalOptics,burkhardtRandomAcceleration}, the saturating self-interacting RW (SATW) \cite{Toth:1996,bremontExactPropagators,Barbier:2020} and short-range correlated processes such as the run-and-tumble particle \cite{solonActiveBrownian}. For RWs outside class I, flips are no longer independent and exhibit memory effects. This long-range memory can either be defined a priori, by imposing slowly decaying correlation functions, or emerge dynamically from the trajectory of the walk itself. The former type is exemplified by Gaussian processes, which are completely determined by the knowledge of the two-point correlation functions. Restricting ourselves to the case of stationary increments, the only such process is fractional Brownian Motion (fBM) \cite{mandelbrotFractionalBrownian}. This defines class (II). For this class, explicit results are given to first order in \(\varepsilon \equiv \dw^{-1} - 1/2\), representing the deviation from Markovian behavior.
The latter type of memory, dynamically generated by the process, is illustrated here by class (III). It is a subclass of non-Gaussian processes and encompasses all non-saturating, self-repelling RWs, which leave behind footprints and interact with them. This class includes the true self-avoiding walk (TSAW), the seminal self-interacting RW \cite{parisi}. \par 
For each class, we have determined the full asymptotics of $\pi_n$ \footnote{For fBM (class (II)), we use $A = e^{-C\varepsilon}$ instead of $A = 1 - C\varepsilon$, ensuring that $A$ remains non-negative. This approach is standard in perturbative theory \cite{wiese}, where both expressions agree to first order in $\varepsilon$.}:
\begin{align}
\label{pin}
    \pi_n \sim \frac{A}{n}, \; A = \begin{cases} \phi  &\text{(I)} \\ 1 -C \varepsilon & \text{(II)} \\ 4/\pi^2 &\text{(III)}, \end{cases}
\end{align}
where $C=4\left(12 \log (\mathcal{G})-\gamma -\frac{7}{3} \log (2) \right) \approx 3.16198$, and $\gamma, \mathcal{G}$ are the Euler and Glaisher constants \cite{abramowitz1965handbook}. 
Our central result \eqref{pin} calls for several comments. (i) The $1/n$ scaling of $\pi_n$ is both strikingly simple and completely universal, holding for all symmetric, asymptotically self-similar RWs. (ii) As outlined in the introduction, the quantity $\pi_n$ is an aged generalization of the classical splitting probability $q_n$, which is defined for a RW starting at the boundary of a \emph{fixed} interval of size $n$. In contrast, $\pi_n$ is defined for a RW at the edge of its dynamically evolving visited territory. The process-dependent scaling $q_n \propto 1/n^\phi$ contrasts sharply with the universal scaling $\pi_n \propto 1/n$. This highlights the substantial and universal impact of memory on the exploration process. Note that $\phi$ can be either larger or smaller than $1$, so that $\pi_n$ can decay more slowly or more rapidly than the non-aged splitting probability (see Fig.~\ref{fig:splitting} below). 
(iii) Lévy walks $J_t$ offer a particularly striking example, as our crossing prescription render the exploration process non-Markovian \footnote{See \eqref{pin-compact} for the arrival convention.}. After visiting—or, in our convention, crossing—its $n^{\text{th}}$ site, $x$, $J_t$ can either stop its jump within the interval $[x, x+1[$, with probability $s_n$, or continue its jump, crossing sites beyond $x+1$. The flip probability is thus given by $\pi_n = s_n q_n + (1-s_n) \times 0$, as $J_t$ cannot flip if it continues its jump. Since $s_n \propto n^{\phi-1}$ \cite{leo}, substituting into $\pi_n = s_n q_n$ reproduces our general scaling relation $\pi_n \propto 1/n$. This inertial-like memory, characteristic of Lévy walks, is fundamentally distinct from the memory present in other non-Markovian RWs, making it all the more remarkable that the $1/n$ scaling holds universally across memory types. 
(iv) The process-dependent prefactor $A$ acts as a fingerprint of the underlying dynamics. It can be determined straightforwardly under the assumption that flips are independent (class (I)), but this naive assumption fails for classes (II) and (III), for which the determination of $A$ requires nontrivial computations, given in SI. In practice, we propose that \(A\) offers a valuable and numerically accessible diagnostic tool for distinguishing memory classes, especially when combined with traditional observables such as mean-square displacement or survival probabilities: \(A\) captures memory effects that cannot be inferred from these classical metrics alone, except in class (I). For instance, the TSAW (class III), and the fBM with \(\dw = \frac{3}{2}\) (class II), both share the same walk dimension \cite{parisi} and persistence exponent \(\theta = \frac{1}{3}\) \cite{bremontPersistenceExponentsa,Barbier:2022,molchan}. Yet, according to \eqref{pin}, their values of $A$ differ by a factor $\approx 1.5$. This underscores the unique information encoded in \(A\) and, more broadly, by flip statistics, as revealed by our approach.

\section*{Numerical and experimental validation}
\begin{figure}[h!]
    \centering
    \begin{minipage}[t]{\columnwidth} 
        \centering
        \begin{subfigure}[t]{0.495\columnwidth}
            \centering
             \includegraphics[width=\textwidth]{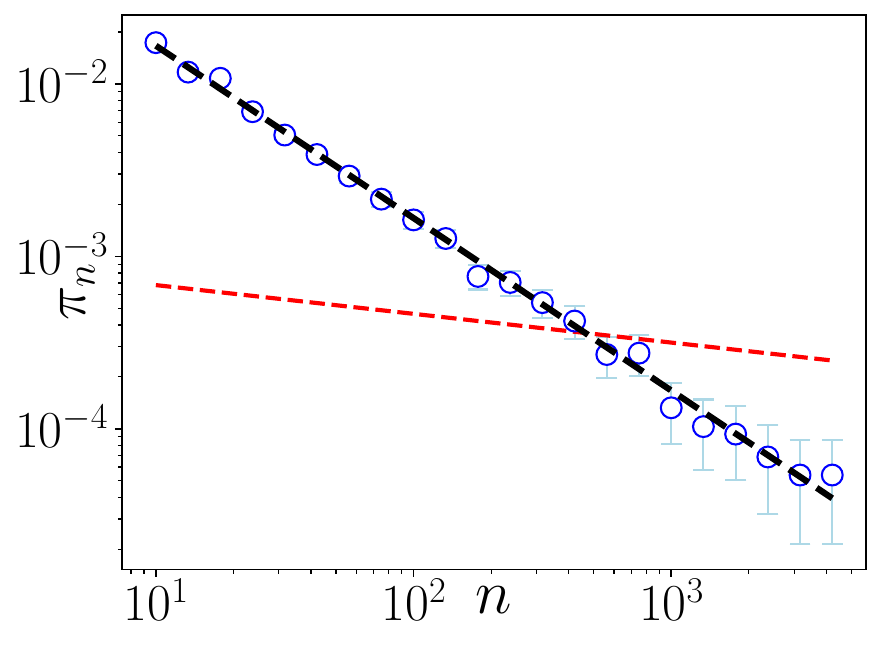}
             \begin{picture}(0,0)
    	\put(-38,30){RAP}
    	\end{picture}
             \label{fig:small1}
        \end{subfigure}%
        \hfill
        \begin{subfigure}[t]{0.495\columnwidth}
            \centering
            \includegraphics[width=\textwidth]{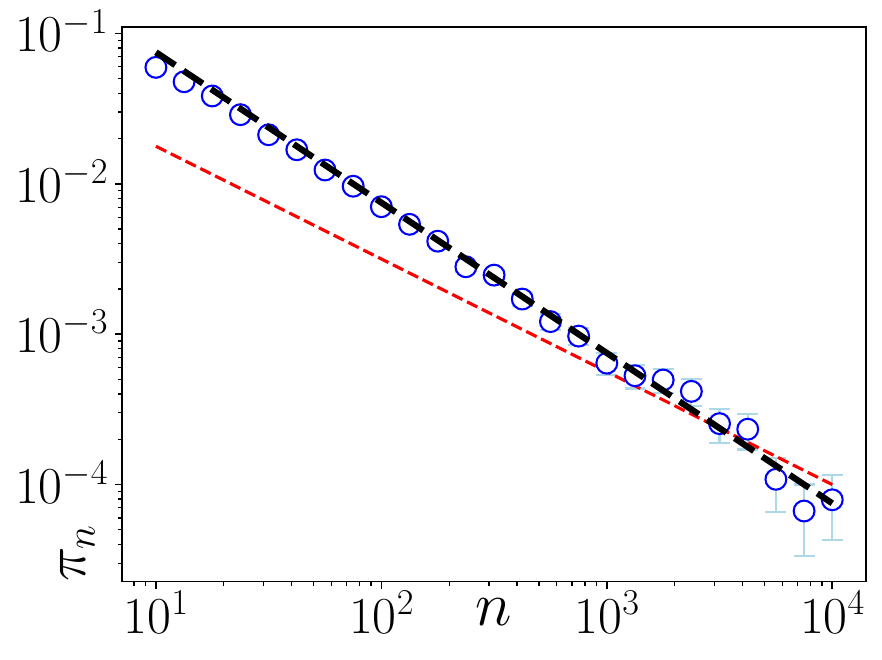}
            \begin{picture}(0,0)
    	\put(-38,30){Lévy Walk, $\beta=1.5$}
    	\end{picture}
            \label{fig:small2}
        \end{subfigure}\\[-1em]
        \begin{subfigure}[t]{0.495\columnwidth}
            \centering
            \includegraphics[width=\textwidth]{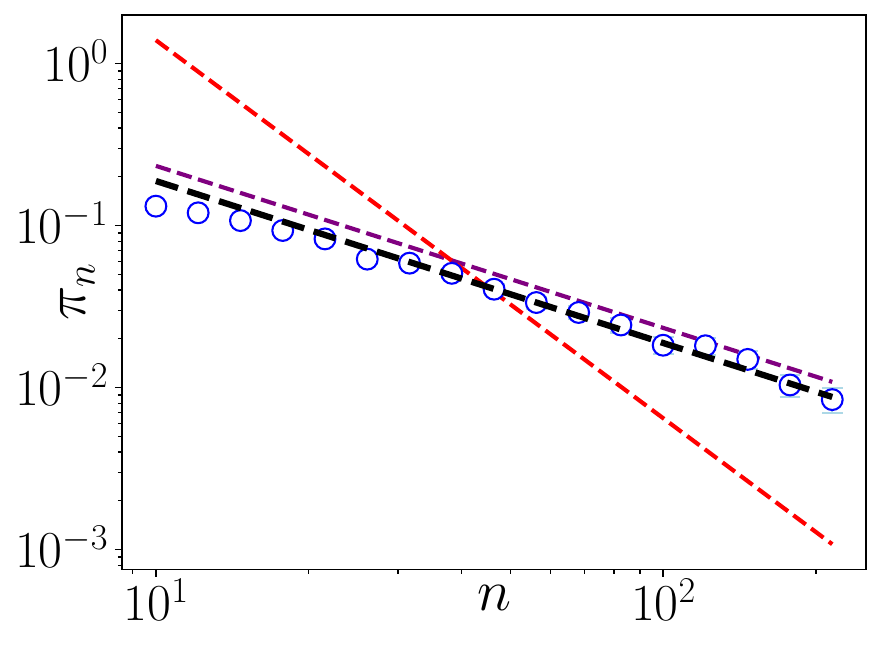}
            \begin{picture}(0,0)
    	\put(-38,30){fBM, $H=0.3$}
    	\end{picture}
            \label{fig:small5}
        \end{subfigure}%
        \hfill
        \begin{subfigure}[t]{0.495\columnwidth}
            \centering
            \includegraphics[width=\textwidth]{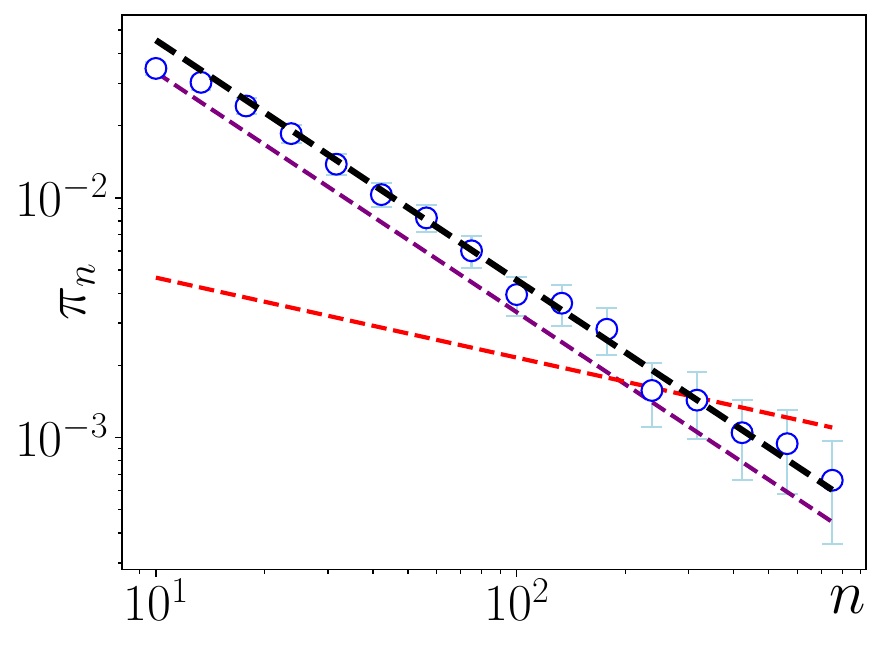}
            \begin{picture}(0,0)
    	\put(-38,30){fBM, $H=0.75$}
    	\end{picture}
            \label{fig:small6}\\[-1em]
        \end{subfigure}
        \begin{subfigure}[t]{0.495\columnwidth}
            \centering
            \includegraphics[width=\textwidth]{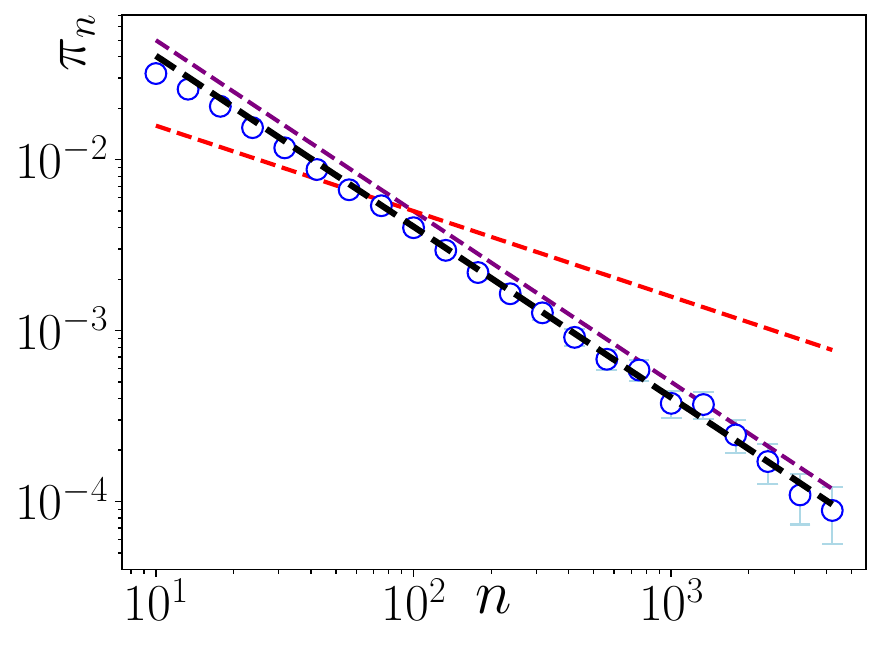}
            \begin{picture}(0,0)
    	\put(-38,30){TSAW}
    	\end{picture}
            \label{fig:small3}
        \end{subfigure}%
        \hfill
        \begin{subfigure}[t]{0.495\columnwidth}
            \centering
            \includegraphics[width=\textwidth]{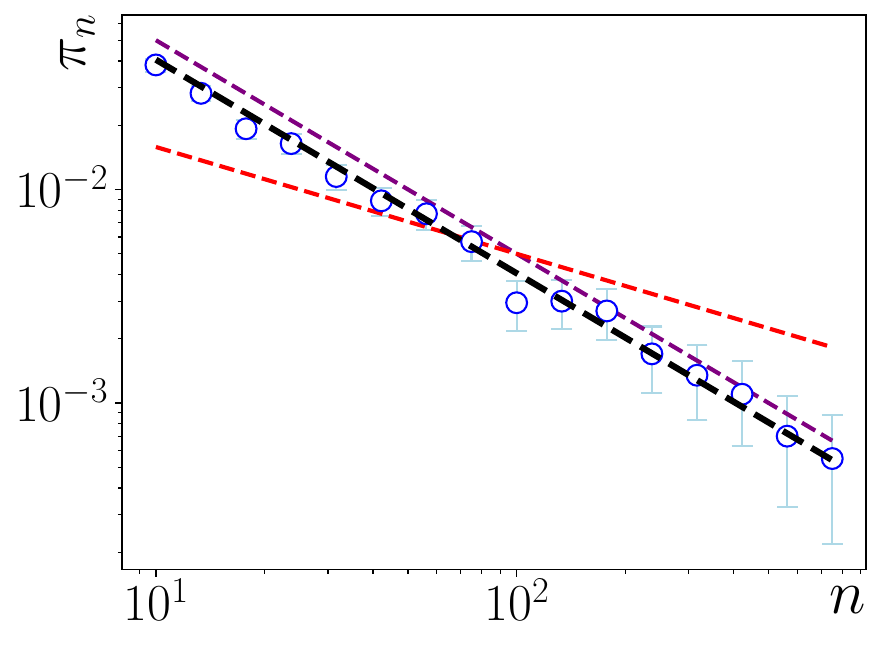}
            \begin{picture}(0,0)
    	\put(-38,30){PSRW}
    	\end{picture}
            \label{fig:small4}
        \end{subfigure}
    \end{minipage}%
    \hfill
    
    \caption{Flip probability $\pi_n$ for representative examples of non-Markovian RWs. Simulation results are in blue. Rows correspond to distinct classes: the first for class (I), the second for class (II), and the third for class (III). The black line shows the theoretical decay $A/n$ with $A$ given by \eqref{pin}, while the red line depicts the classical splitting probability, decaying as $1/n^\phi$. The purple dashed line illustrates the naive estimate $\pi_n \sim \frac{\phi}{n} = \frac{\dw \theta}{n}$, which holds only when flips are uncorrelated (class (I)). Outside of class (I), this approximation either overestimates the flip probability (as in class (III) or antipersistent fBM, class (II)) or underestimates it (as in persistent fBM, class (II)). Errors in the prefactor $A$ are below \(0.1\%\) for the exact results of classes (I) and (III), and around \(3.5\%\) for the perturbative result of class (II).
    }
    \label{fig:splitting}
\end{figure}
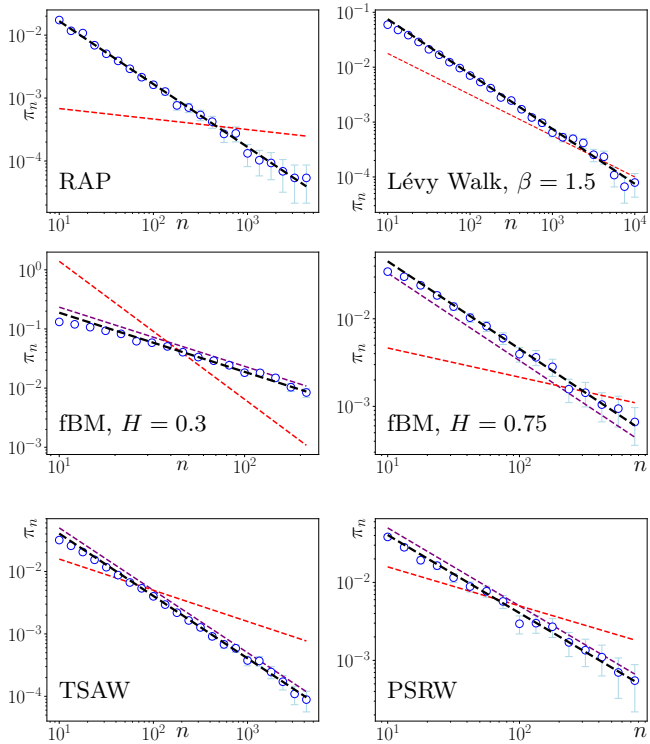

We now validate our theoretical prediction \eqref{pin} on paradigmatic non-Markovian models, which illustrate classes (I)-(III) defined above and are defined in the Methods section. Overall, we find excellent quantitative agreement between our theoretical predictions and numerical simulations (see Fig.~\ref{fig:splitting}). The diversity of the models demonstrate the robustness and universality of our theoretical predictions, further highlighting the practical relevance of our findings in various scientific domains. 

We now demonstrate the experimental relevance of our results by analyzing effectively one-dimensional datasets of very diverse origins. Crucially, the $1/n$ scaling law holds even when the specific mechanisms responsible for non-Markovian behavior are unknown, as is the case of most real-world observations (see the detailed analysis in the Methods section). We begin with one-dimensional projections of biological tracers at different scales, which serve as natural realizations of the exploration setting. (a) and (b) concern displacements in vitro of single human cells ~\cite{Phillip:2021}; (c) displays intracellular motion of amoeba vacuoles~\cite{Krapf:2019}; and (d) depicts telomere dynamics within the nucleus~\cite{Stadler:2017}. We also analyze time series that exhibit long-range memory, such as (e) and (f) DNA sequences \footnote{DNA sequences are Pyrimidines/Purines HUMBMYH7 and HUMTCRADCV.}, represented as RWs where each step is \(+1\) for guanine (G) and \(-1\) for cytosine (C) \cite{GenBank}. This mapping reveals key features of genomic organization, such as GC skew, which highlights replication origins and termini \cite{gc1, gc2, gc3}. Flips in the RW trajectory mark transitions between G-rich and C-rich regions, offering a model-free indicator of genomic asymmetry. 
Finally, (g), (h) and (i) show financial indices SPX, N225 and Dow-Jones, where flips correspond to market drawdown events— declines in the market from a recent peak to a subsequent trough \cite{rejYouAre}— and corrections. \par
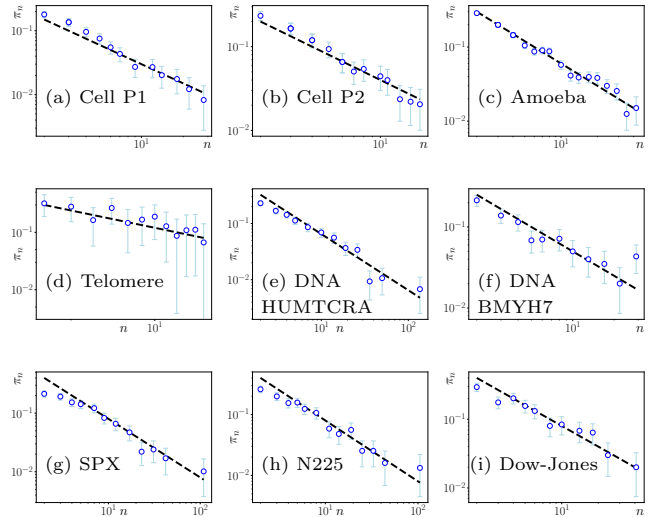
\begin{figure}
    \begin{subfigure}{0.32\columnwidth}
		\includegraphics[width=\textwidth]{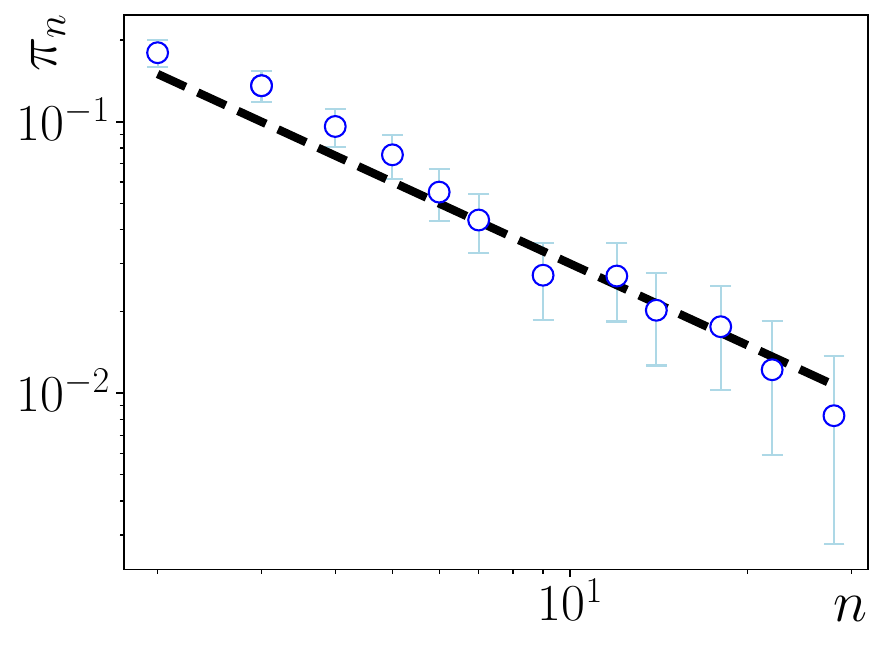}
		\begin{picture}(0,0)
    	\put(-25,30){\scriptsize (a) Cell P1}
    	\end{picture}
    \end{subfigure}
    \begin{subfigure}{0.32\columnwidth}
		\includegraphics[width=\textwidth]{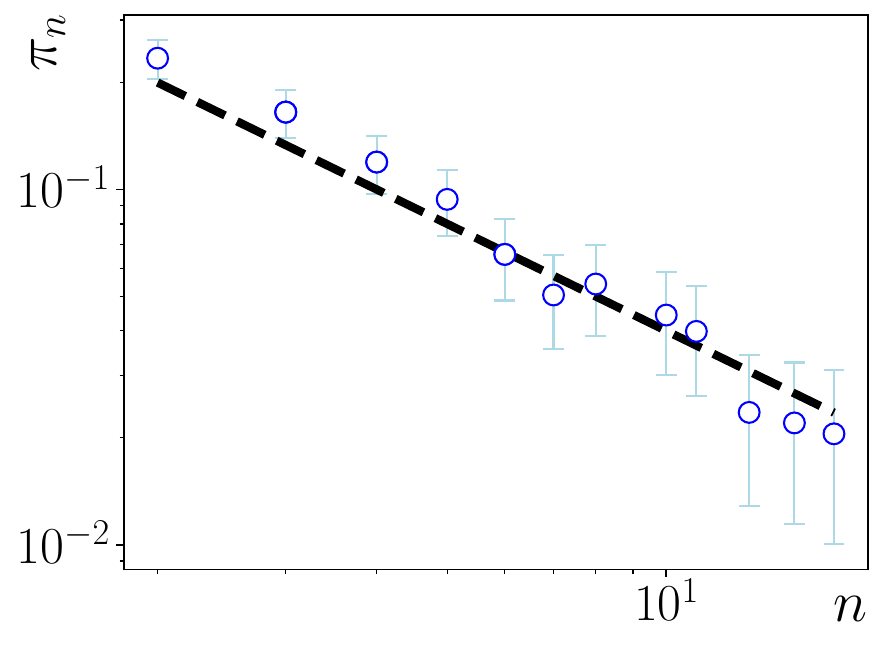}
		\begin{picture}(0,0)
    	\put(-25,30){\scriptsize (b) Cell P2}
    	\end{picture}
    \end{subfigure}
    \begin{subfigure}{0.32\columnwidth}
		\includegraphics[width=\textwidth]{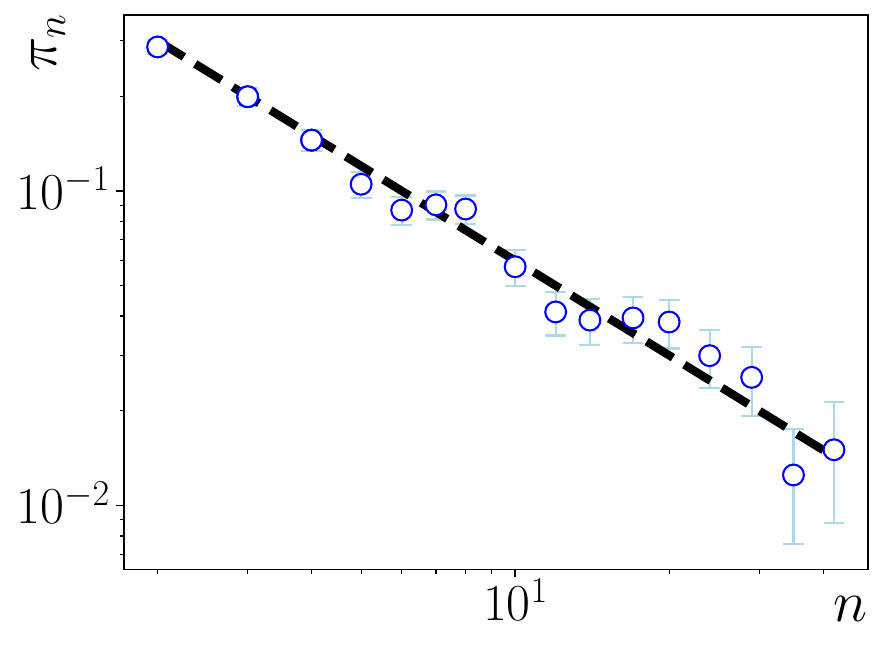}
		\begin{picture}(0,0)
    	\put(-25,30){\scriptsize (c) Amoeba}
    	\end{picture}
    \end{subfigure}
    \begin{subfigure}{0.32\columnwidth}
      \includegraphics[width=\textwidth]{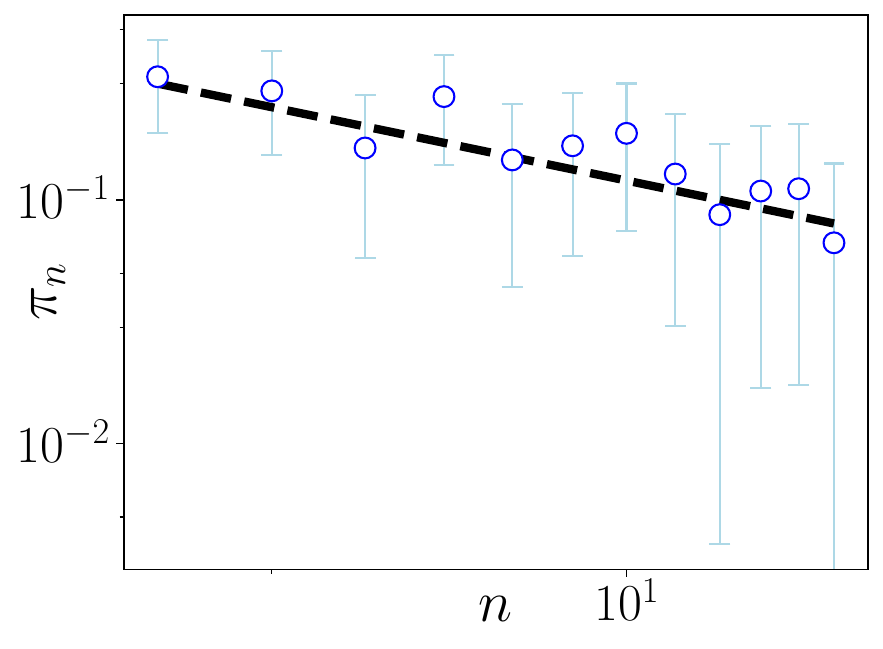}
      \begin{picture}(0,0)
    	\put(-25,30){\scriptsize (d) Telomere}
    	\end{picture}
    \end{subfigure}
    \begin{subfigure}{0.32\columnwidth}
		\includegraphics[width=\textwidth]{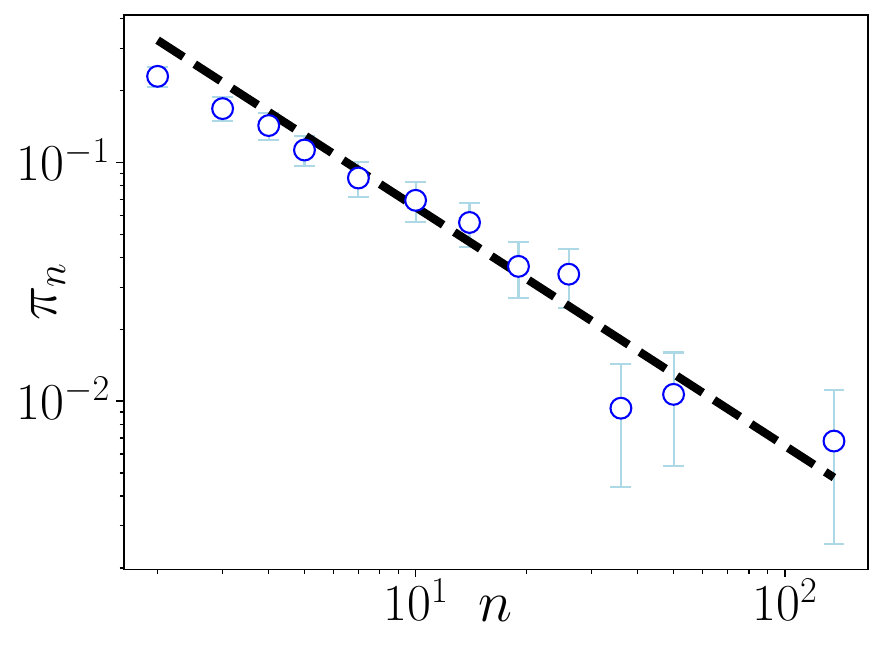}
		\begin{picture}(0,0)
    	\put(-25,30){\scriptsize (e) DNA}
    	\put(-25,20){\scriptsize HUMTCRA}
    	\end{picture}
    \end{subfigure}
    \begin{subfigure}{0.32\columnwidth}
		\includegraphics[width=\textwidth]{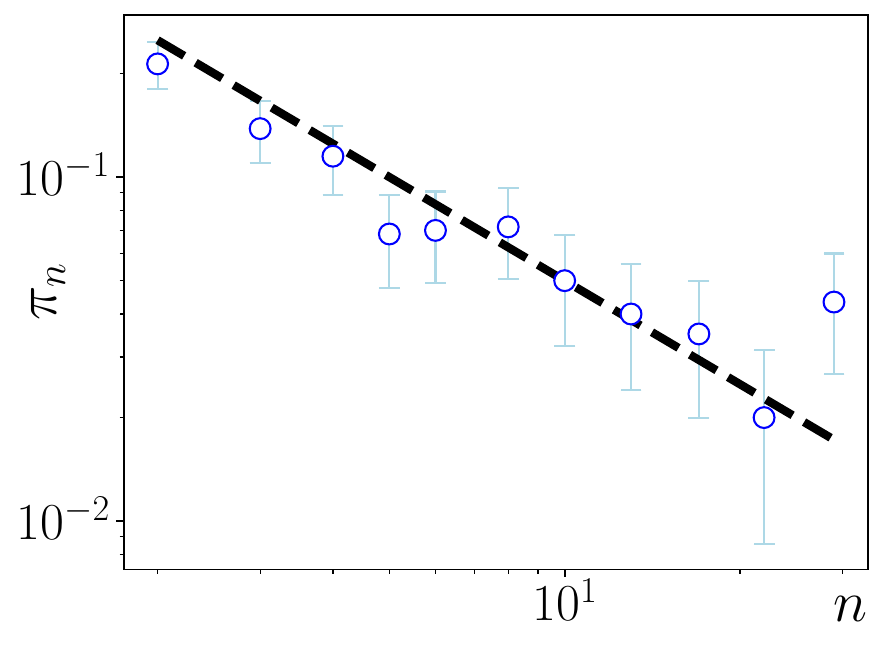}
		\begin{picture}(0,0)
    	\put(-25,30){\scriptsize (f) DNA}
    	\put(-25,20){\scriptsize BMYH7}
    	\end{picture}
    \end{subfigure}
    \begin{subfigure}{0.32\columnwidth}
		\includegraphics[width=\textwidth]{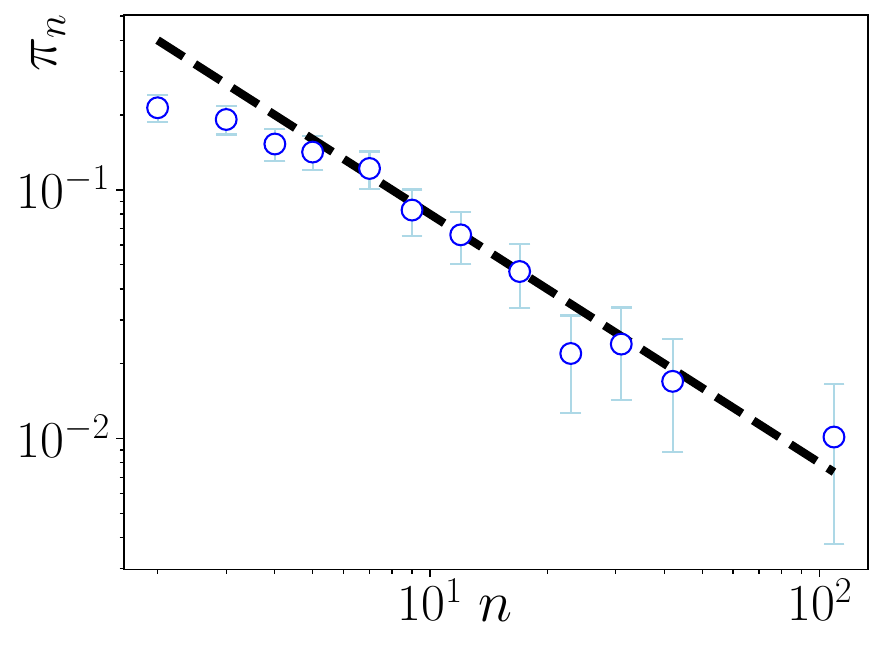}
		\begin{picture}(0,0)
    	\put(-25,30){\scriptsize (g) SPX}
    	\end{picture}
    \end{subfigure}
    \begin{subfigure}{0.32\columnwidth}
		\includegraphics[width=\textwidth]{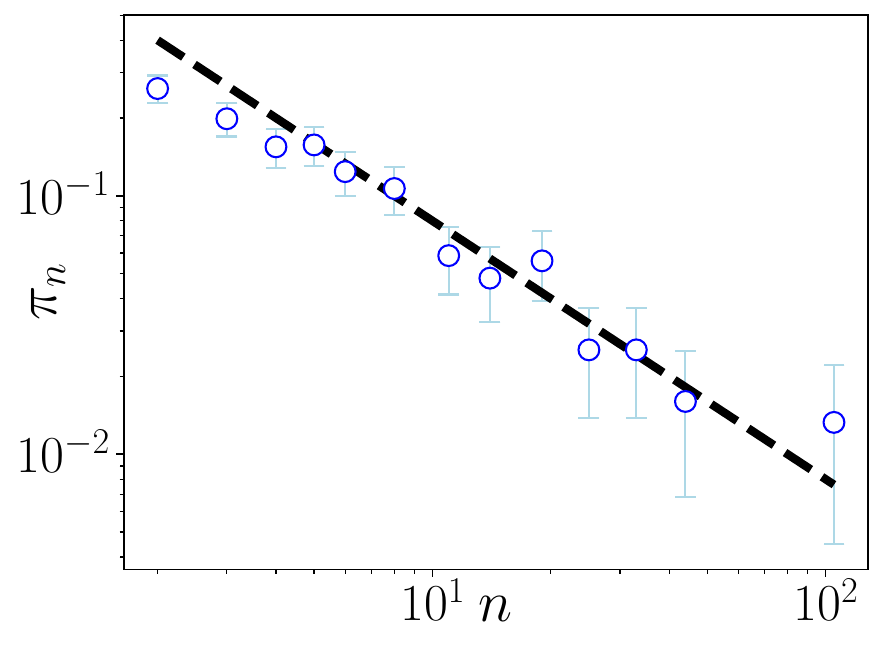}
		\begin{picture}(0,0)
    	\put(-25,30){\scriptsize (h) N225}
    	\end{picture}
    \end{subfigure}
    \begin{subfigure}{0.32\columnwidth}
      \includegraphics[width=\textwidth]{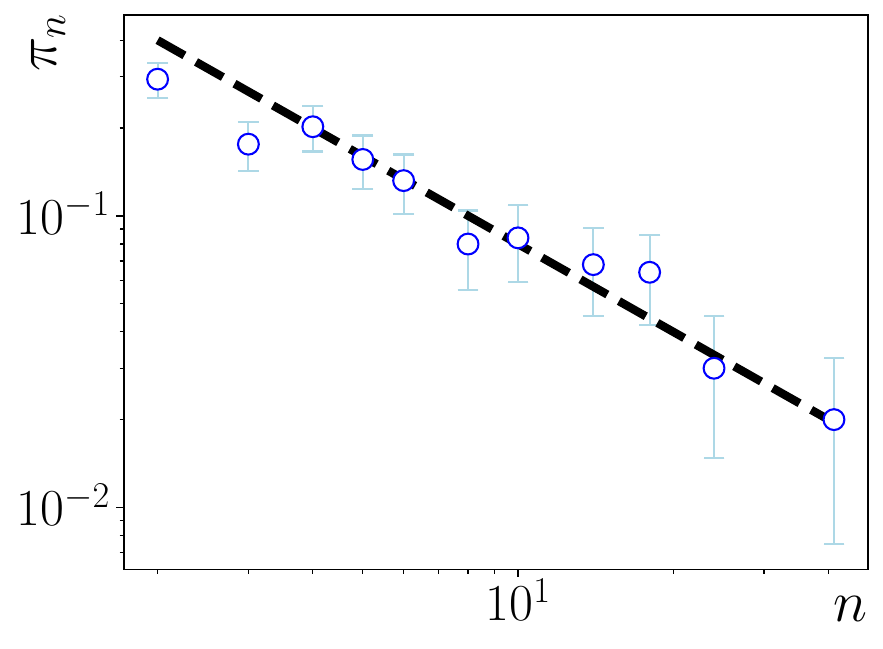}
      \begin{picture}(0,0)
    	\put(-28,30){\scriptsize (i) Dow-Jones}
    	\end{picture}
    \end{subfigure}
    
    \caption{Flip probability $\pi_n$ is shown for each dataset (blue symbols). The black line indicates the universal $1/n$ scaling. Panels (a) and (b) refer to single-cell  (dermal fibroblasts) displacements in vitro from young and older patients  ~\cite{Phillip:2021}; (c) presents intracellular motion of amoeba vacuoles~\cite{Krapf:2019}; and (d) shows telomere dynamics within the nucleus~\cite{Stadler:2017}. Panels (e) and (f) correspond to DNA sequences represented as RWs, where each step is \(+1\) for guanine (G) and \(-1\) for cytosine (C). Finally, (g), (h), and (i) are financial indices SPX, N225, and Dow Jones, respectively.
    }
    \label{fig:pi_n_real}
\end{figure}

 
Fig.~\ref{fig:pi_n_real} demonstrates the quantitative agreement between these diverse real-world datasets and our $1/n$ scaling law. Let us illustrate the practical consequences of our analysis by focusing on the example of financial markets. Our main result \eqref{pin} implies that the probability of a drawdown event decays inversely with the market’s exploration history (i.e., the number of distinct price levels visited \footnote{Note that after the logarithmic transform explained in the Methods section, $n$ cannot be interpreted as the size of the visited interval, but is indeed the number of distinct log-indices visited.}). Remarkably, all three financial indices SPX, N225 and Dow-Jones exhibit a similar prefactor, \(A \approx 0.8\), pointing to a common underlying mechanism across markets that is not captured by standard volatility-based metrics, which often fail to account for memory effects \cite{vanbuskirkVolatilitySkew}. While this precise mechanism is not clear, it is plausible that such behavior reflects a common response to global economic conditions. Flips thus reveal, and may act as a statistical signature of, an emergent structure in the statistical patterns of market dynamics—one that transcends individual market features, correlates them, and points to a fundamental, system-wide behavior. \par
Finally, flips offer a simple and model-free tool for probing memory-driven exploration in real-world systems. As universal statistical features, the $1/n$ scaling law of flips does not reflect specific underlying mechanisms. Instead, flips offer a robust and predictive metric for reversal events that can be directly extracted from the knowledge of the number $n$ of past levels of raw time series, without requiring knowledge of the system’s internal dynamics, and serve as a valuable complement to traditional measures such as mean-square displacement or survival probabilities. In addition to the universal \(1/n\) scaling, the (measurable) prefactor \(A\) in the flip probability \(\pi_n = A/n\) encodes system-specific memory effects and quantifies the system's propensity to reverse its direction of exploration. 

\begin{figure}
  \centering
  \includegraphics[width=.8\columnwidth]{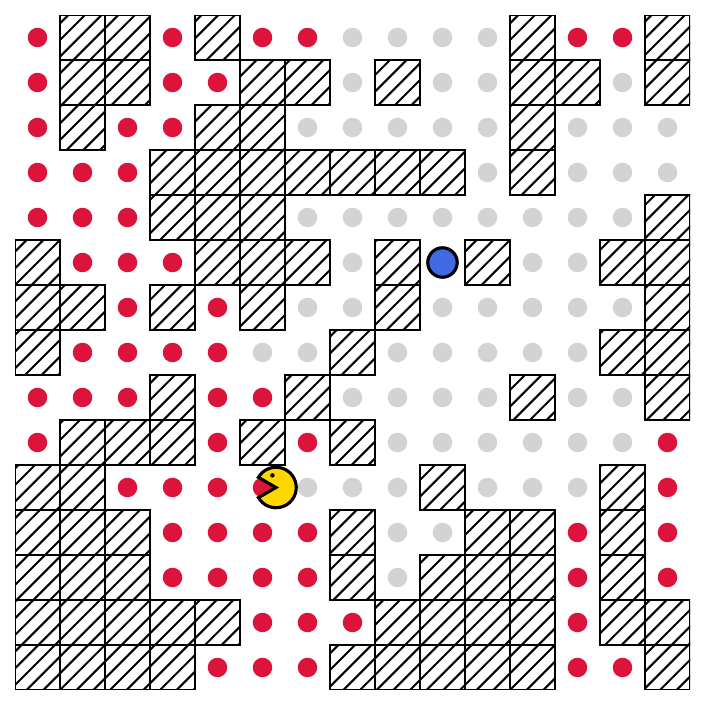}
  \caption{Illustration of flips in a higher-dimensional medium. This example shows a RW on a critical percolation cluster. Grey balls indicate sites that have already been visited, while red balls mark unexplored sites. The large blue ball represents the origin of the walk, which acts as an infinite reservoir of blue balls. The walker (in yellow) has just consumed a red ball. Similarly to the one-dimensional case, a flip occurs if the walker consumes a blue ball before eating another red one.}        
  \label{fig:Illus2}
\end{figure}
\section*{Flips in higher dimensions}
Importantly, we underline that our findings are not limited to the one-dimensional setting, and extend to all compact, symmetric and asymptotically scale-invariant RWs. Such RWs eventually come back to their starting site with probability $1$. They include the physically important cases of RWs on deterministic or random fractals \cite{bancaud}, such as critical percolation clusters, which is a minimal model of transport in a crowded environment \cite{bancaud2}. In this extended setting, a flip occurs when the walker, after visiting the \(n\)th new site, returns to the origin before discovering the \((n{+}1)\)st (see Fig.~\ref{fig:Illus2}). This mimics a foraging scenario, where the origin acts as a nest and flips mark returns home between outward excursions. While differing in form from the original $1d$ definition, this notion preserves its essence: flips are excursions into previously explored territory before further expansion. In fact, our derivation below shows in particular that for $d = 1$, both definitions lead to the same $1/n$ scaling law, albeit with different prefactors (see SI).
Notably, for $d>1$, flips remain history-dependent even for Markovian RWs. This is because, although the number of visited sites is fixed to $n$, the geometry of the explored region remains random, see Fig.~\ref{fig:Illus2}. Strikingly, we find that our scaling result \eqref{pin} still holds in this extended setting. Indeed, for compact RWs, the distribution of the time \( \tau_n \) between the \( n \)th and \( (n+1) \)st site discoveries has recently been shown to follow the asymptotic form \cite{regnierUniversalExplorationDynamics2023}:
\begin{equation}
    \label{leo}
    \mathbb{P}(\tau_n = \tau) \sim \frac{1}{n^{1 + 1/\mu}} \, \psi\left( \frac{\tau}{n^{1/\mu}} \right),
\end{equation}
which amounts to stating that the visited territory is scale invariant for compact RWs. In \eqref{leo}, \( \mu \equiv \frac{\df}{\dw} < 1\), $\psi$ is a process-dependent scaling function, and \( \df \) is the fractal dimension of the medium. The visited region $V_n$ is essentially hole-free for compact RWs, and spans a linear size \( R \propto n^{1/\df} \). A flip occurs at the $(n+1)$st visit if the walker, starting at the boundary \( \partial V_n \), returns to the origin before crossing \( \partial V_n \). This requires that the discovery time \( \tau_n \) be larger than the typical time to reach the origin from the $n$th visited site, i.e., \( \tau_n \gtrsim R^{d_w} = n^{1/\mu} \). After this time, the walker returns to the origin before crossing the boundary with probability $p_n^0 \sim 1$ for compact RWs. Therefore, the flip probability scales as:
\begin{equation}
\label{pin-compact}
        \pi_n^{\text{compact}} \propto \left(\int_{n^{1/\mu}}^\infty \mathbb{P}(\tau_n = \tau) \, d\tau \right) \times p_n^0 \propto \frac{1}{n},
\end{equation}
with the last $\propto$ sign being a consequence of \eqref{leo}. In turn, this provides the physical mechanism responsible for the universal $1/n$ behavior.

\begin{figure}[h!]
    \centering
    \begin{minipage}[t]{\columnwidth} 
        \centering
        \begin{subfigure}[t]{0.32\columnwidth}
            \centering
             \includegraphics[width=\textwidth]{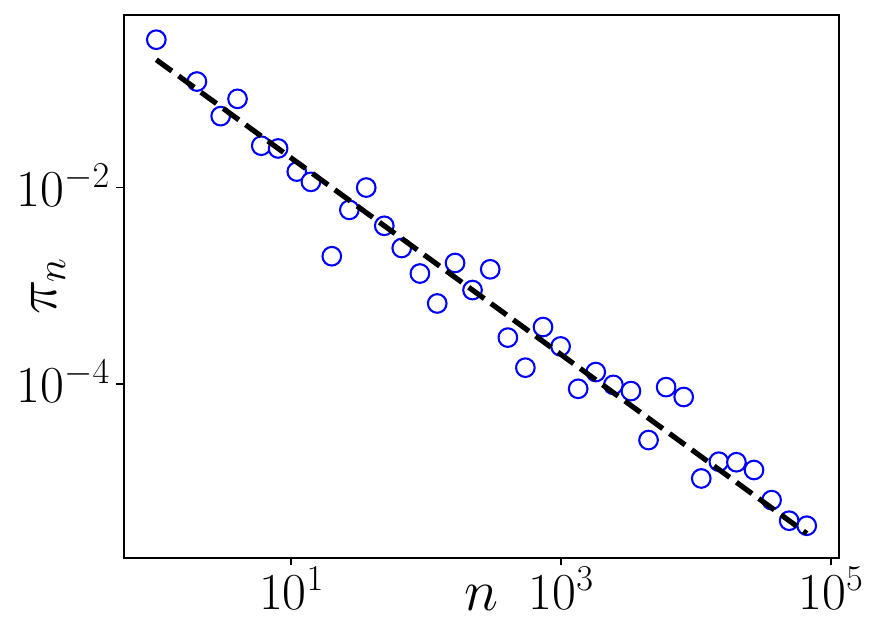}
             \begin{picture}(0,0)
    	\put(-15,60){\scriptsize Sierpiński RW}
    	\end{picture}
        \end{subfigure}%
        \hfill
        \begin{subfigure}[t]{0.32\columnwidth}
            \centering
            \includegraphics[width=\textwidth]{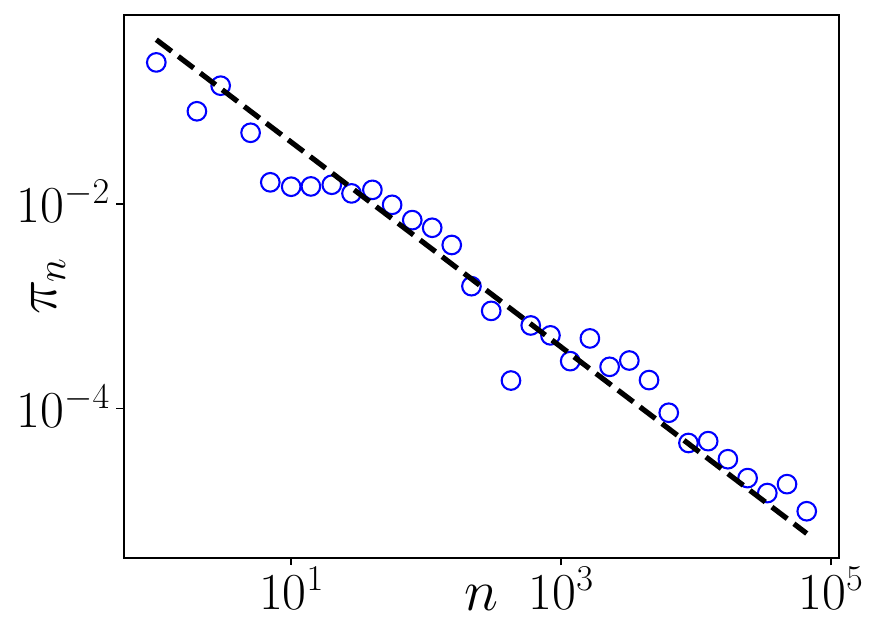}
            \begin{picture}(0,0)
    	\put(-20,60){\scriptsize Percolation RW}
    	\end{picture}
        \end{subfigure}
        \begin{subfigure}[t]{0.32\columnwidth}
            \centering
            \includegraphics[width=\textwidth]{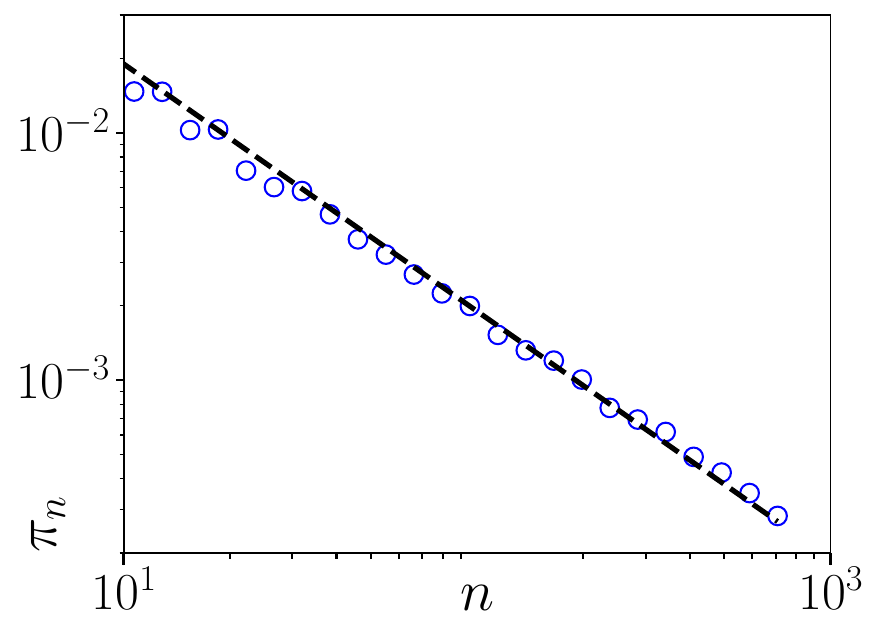}
            \begin{picture}(0,0)
    	\put(-10,60){\fontsize{6pt}{10pt}\selectfont Lévy Walk}
        \put(-10,55){\fontsize{6pt}{10pt}\selectfont $\beta=1.5$}
    	\end{picture}
        \end{subfigure}
        \\[-1em]
        \begin{subfigure}[t]{0.48\columnwidth}
            \centering
            \includegraphics[width=\textwidth]{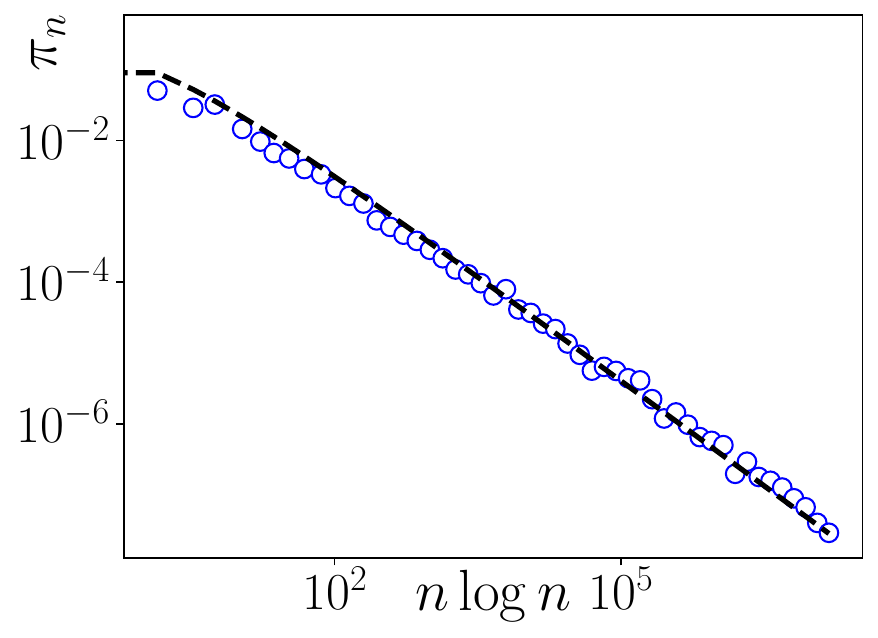}
            \begin{picture}(0,0)
    	\put(-15,40){2d RW}
    	\end{picture}
        \end{subfigure}\hfill 
        \begin{subfigure}[t]{0.48\columnwidth}
            \centering
            \includegraphics[width=\textwidth]{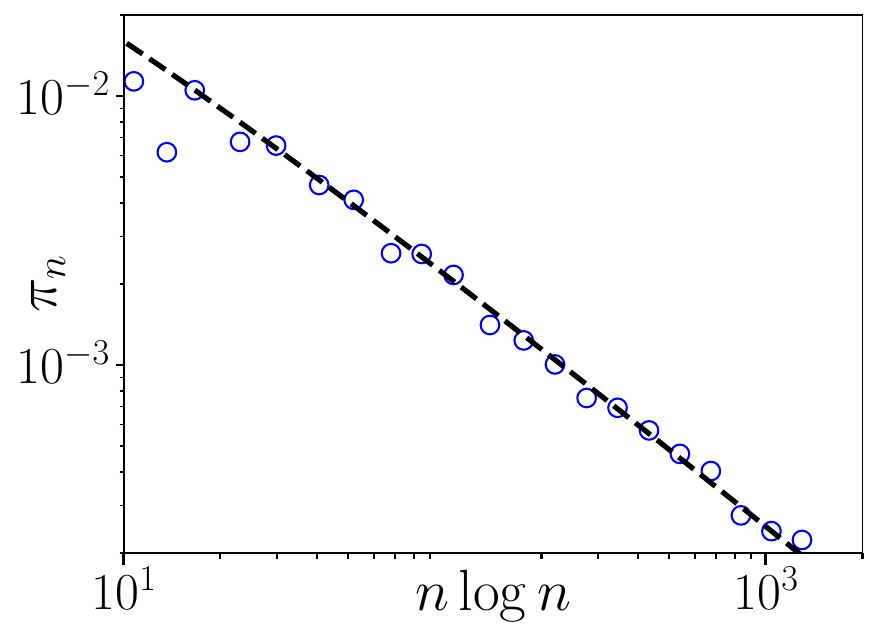}
            \begin{picture}(0,0)
    	\put(-35,40){Lévy Walk}
        \put(-35,30){ $\beta=1$}
    	\end{picture}
        \end{subfigure}
         \\[-1em]
         \begin{subfigure}[t]{0.32\columnwidth}
            \centering
             \includegraphics[width=\textwidth]{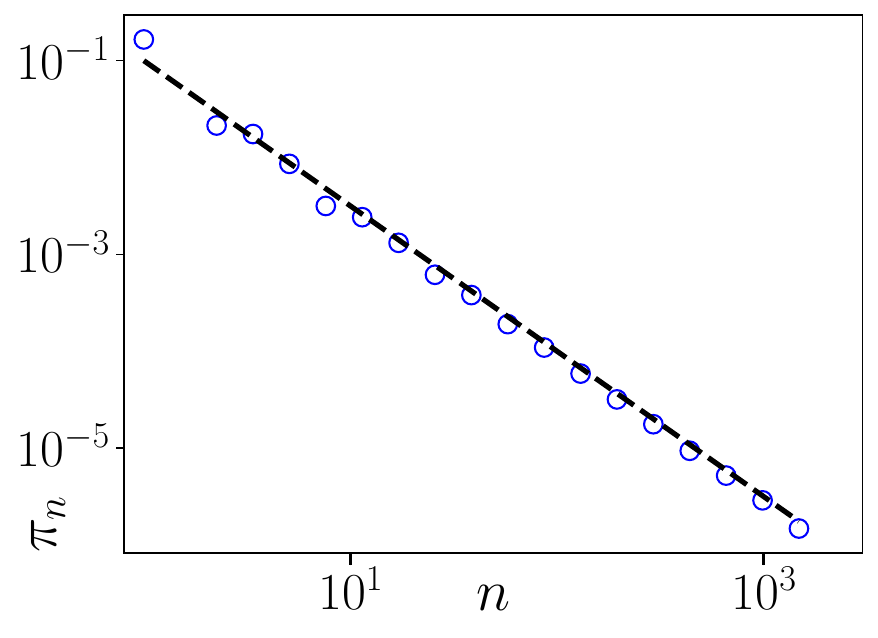}
             \begin{picture}(0,0)
    	\put(-10,60){\scriptsize 3d RW}
    	\end{picture}
        \end{subfigure}%
        \hfill
        \begin{subfigure}[t]{0.32\columnwidth}
            \centering
            \includegraphics[width=\textwidth]{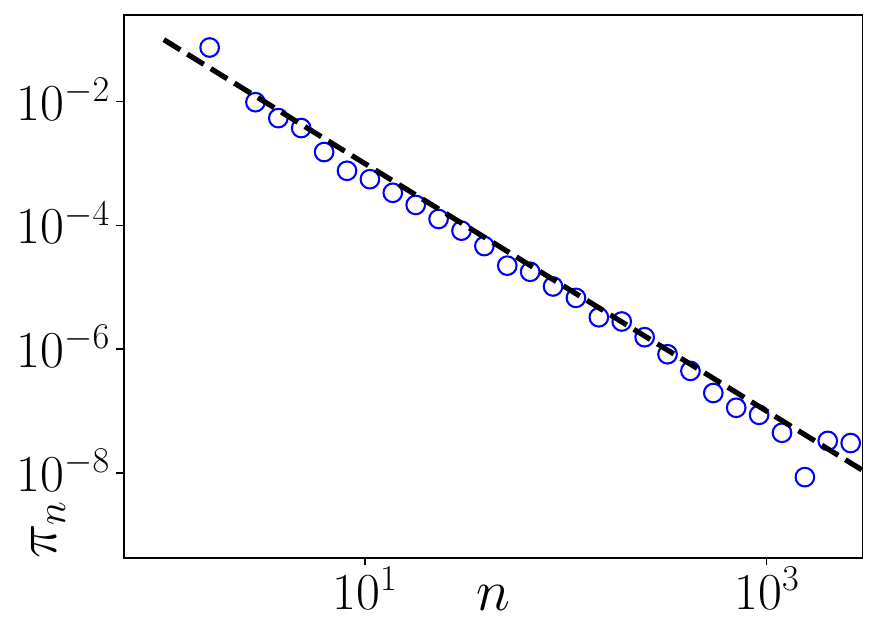}
            \begin{picture}(0,0)
    	\put(-10,60){\scriptsize 4d RW}
    	\end{picture}
        \end{subfigure}
        \begin{subfigure}[t]{0.32\columnwidth}
            \centering
            \includegraphics[width=\textwidth]{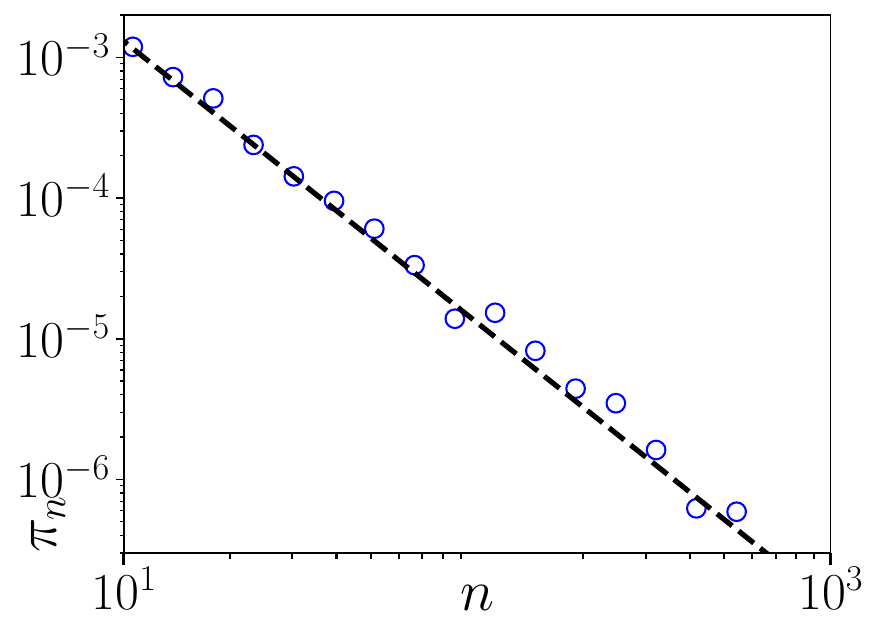}
            \begin{picture}(0,0)
    	\put(-10,60){\fontsize{6pt}{10pt}\selectfont Lévy Walk}
        \put(-10,55){\fontsize{6pt}{10pt}\selectfont $\beta=0.5$}
    	\end{picture}
        \end{subfigure}
    \end{minipage}%
    \hfill
    
    \caption{Scaling of the flip probability $\pi_n$ in dimension $\df>1$, and for 1d Lévy walks with the arrival prescription, where long jumps leave holes in the visited domain. Simulation results are in blue. The first row shows compact RWs, including walks on deterministic fractals (e.g., the Sierpiński gasket), random fractals (e.g., critical percolation clusters), and one-dimensional compact Lévy Walks with arrival convention ($\beta > 1$). For all these, the flip probability follows the universal scaling $\pi_n \propto 1/n$ \eqref{pin-compact}. The second row illustrates marginally recurrent walks—the 2D simple RW and the critical $1d$ Lévy Walk with $\beta = 1$—which exhibit slower decay, $\pi_n \gtrsim 1/(n \log n)$ \eqref{pin-marginal}. The third row corresponds to transient walks, including 3D and 4D simple RWs and transient $1d$ Lévy Walks ($\beta < 1$), for which the flip probability decays as $\pi_n \gtrsim 1/n^{\mu}$ \eqref{pin-transient}.}
    \label{fig:splitting-highd}
\end{figure}
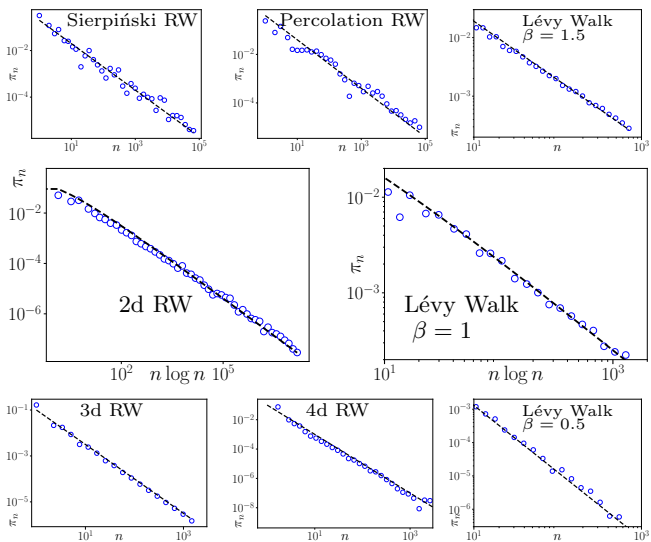

In marginally recurrent RWs, exemplified by the $2d$ simple RW, the $1/n$ scaling holds up to logarithmic corrections. In the SI, we derive the lower bound: 
\begin{equation}
\label{pin-marginal}
    \pi_n^{\text{marginal}} \gtrsim
     \frac{1}{n \log n}.
\end{equation}
Finally, in non-compact, or transient, cases, such as the $3d$ simple RW, it is striking that the flip probability still exhibits algebraic decay:
\begin{equation}
\label{pin-transient}
    \pi_n^{\text{transient}} \gtrsim
     \frac{1}{n^{\mu}}, \quad \mu=\frac{\df}{\dw}>1.
\end{equation}
The universal scaling law \eqref{pin-compact} and the accuracy of the lower bounds \eqref{pin-marginal}, \eqref{pin-transient} are confirmed numerically Fig.~\ref{fig:splitting-highd}. \par 

\section*{Discussion and outlook}
In conclusion, we have uncovered a universal scaling law for the probability \(\pi_n\) of a flip in non-Markovian RWs: \(\pi_n \propto 1/n\). This simple yet unexpected result holds across a wide range of stochastic processes and real-world systems—from intracellular transport to genomic sequences and financial markets—highlighting the fundamental role of memory in shaping exploration.
Our findings overturn conventional expectations: memory, often associated with complexity and model-specific behavior, instead gives rise to a universal structure governing the dynamics of exploration. This leads to two central insights. First, flips constitute a fundamental statistical feature that transcends microscopic details of the underlying dynamics. Second, conditioning on the number of distinct sites visited—rather than on physical time—induces universal behavior in memory-driven processes. While traditional observables in non-Markovian dynamics depend intricately on the full trajectory history, reframing the problem in terms of exploration history simplifies the picture dramatically. The number of visited sites emerges as a natural “clock” for exploration, one that effectively absorbs memory effects and reveals a deeper geometric invariance. Beyond providing a new observable with universal statistics, our results open a broader perspective, by identifying flips as the fundamental units of memory-driven exploration.
\section*{Methods}

\subsection*{Definitions of paradigmatic non-Markovian processes}

(I-i) \textbf{Lévy walks} \cite{klafterFirstSteps,vezzaniFastRare} are jump processes characterized by step lengths \( l \) drawn from a power-law distribution \( p(l) \propto l^{-\beta - 1} \), with \( 0 < \beta < 2 \). This fat-tailed behavior induces non-Markovian statistics, observed in various natural systems including bacterial motion~\cite{bacteria} and human mobility patterns~\cite{human}.

(I-ii) \textbf{Random Acceleration Process (RAP)} is one of the simplest non-Markovian processes \cite{burkhardtRandomAcceleration}. It corresponds to integrated Brownian motion, \( X_t = \int_0^t B_u \, \mathrm{d}u \), where the position results from integrating a Brownian velocity. This model is fundamental in describing systems where momentum or inertia leads to long-range speed correlations.

(II) \textbf{Fractional Brownian Motion (fBM)}~\cite{Mandelbrot:1968} is a Gaussian process with stationary increments obeying \( \langle (X_t - X_s)^2 \rangle = |t - s|^{2H} \), where \( 0 < H = d_w^{-1} < 1 \) is the Hurst exponent. As a representative of class~(II), fBM captures long-range temporal correlations and has been applied to anomalous diffusion, seismic activity~\cite{earthquake}, viscoelastic transport~\cite{Krapf:2019}, and financial time series~\cite{volatility}.

(III) \textbf{TSAW and Polynomial Self-Repelling Walk (PSRW)} are key examples of self-interacting RWs~\cite{Toth:1995,Barbier:2022,parisi,hokmabadChemotacticSelfcaging}. Here, the walker’s transition probabilities depend explicitly on its visitation history, typically through the number of previous visits to neighboring sites (see SI for precise definitions). Such self-interactions are not merely theoretical constructs—they have been observed in diverse real-world contexts, from cellular motion~\cite{cellattract,naturealex,lucas} to non-reversible Monte Carlo algorithms~\cite{maggs1}, and animal foraging~\cite{ants,animals}.

\subsection*{Data analysis: Windowing method}

Our theoretical framework predicts a universal \( 1/n \) scaling law for flip probabilities in symmetric, scale-invariant processes. However, experimental data usually consist of a single trajectory, precluding ensemble averaging.

To address this, we employ a simple and general windowing method~\cite{Regnier:2023}. A trajectory \( X_t \) of duration \( T \) is divided into \( N \) overlapping sub-trajectories. Each sub-trajectory \( k \) begins at time \( kT/N \) and extends to the final time \( T \). Flips are then defined with respect to the dynamically explored region within each sub-trajectory—treating each window as an effective realization of the process. This approach enables the robust extraction of flip statistics from a single trajectory.

Notably, flips identified via this method correspond to the alternative definition illustrated in Fig.~\ref{fig:Illus2}, which captures a broader class of reversals (e.g., returns to a reference point, like the origin, before visiting a new site). Remarkably, we observe that the universal \( 1/n \) scaling persists in such single-trajectory data across a wide range of systems.

A particularly illustrative example is provided by financial indices such as the SPX, N225, and Dow-Jones. These processes are not strictly scale-invariant and would therefore appear to lie outside the direct scope of our scaling result Eq. \eqref{pin}. However, it is well established in the financial literature~\cite{bouchaudTheoryFinancial} that such indices typically follow an exponential growth trend—similar to geometric Brownian motion, \( Y_t = e^{B_t} \). While the process \( Y_t \) is not scale-invariant, its logarithm \( \log Y_t \) is. Accordingly, we analyze the scale-invariant log-indices \( \log X_t \) rather than the financial index \( X_t \) itself. Since the logarithm is a strictly increasing function, flips are preserved, while the exponential growth is effectively removed. Despite the presence of a small residual drift (\( \sim 10^{-5} \)–\( 10^{-4} \) per time step), the \( 1/n \) scaling remains robust, illustrating the relevance of our framework even in weakly biased systems.

\section*{Acknowledgments}
R.V. acknowledges support of ERC synergy grant SHAPINCELLFATE.

\section*{Contributions}
J.B. carried out the analytical calculations. J.B. and L.R. conducted the numerical simulations and analyzed the experimental data. J.B., O.B. and R.V. jointly wrote the manuscript. The idea of flips originated from the work of A.B.–C. O.B. and R.V. conceived the research.

\clearpage
\appendix
\onecolumngrid
\section*{Supplementary Material}
\section{Description of non-Markovian models}
In this section, we describe each of the numerical models used in the main text.
\begin{itemize}
    \item \emph{Fractional Brownian motion} (fBm) \cite{Mandelbrot:1968,Hansen:1994,Ding:1995,Maslov:1994} is a non-Markovian Gaussian process, with stationary increments. Thus, an fBm $X_t$ of Hurst index $H$ is defined by its covariance 
\begin{align}
    \text{Cov}\left(X_t,X_{t'} \right)= \frac{1}{2} \left(t^{2H} + t'^{2H}-|t-t'|^{2H} \right) \; .
\end{align}
    The steps $\eta_t=X_t-X_{t-1}$ are called fractional Gaussian noise (fGn). 
    Nowadays, the fBm is broadly spread and its implementations could be found in standard packages of python or Wolfram Mathematica.
    \item \emph{RAP.} the Random Acceleration Process (RAP) is defined as the integrated Brownian motion: if $B_t$ is a BM, then 
    \begin{align}
        X_t=\int_0^t B_{t'} {\rm d}t'
    \end{align}
    is a RAP.
    \item \emph{L\'evy Walk.} A L\'evy walk of parameter $0<\alpha<2$ is a RW model where successive jumps of length $\ell$ are drawn from a Pareto law, $p(\ell)\propto \frac{1}{\ell^{1+\alpha}}$ \cite{Palyulin:2019}. The direction is then either positive or negative with equal probability, and the RW moves ballistically along that direction for a length $\ell$. Importantly, the jumps are not instantaneous, and the RW moves with speed $1$. The combination of (i) finite speed and (ii) the non-exponential distribution of jump lengths make the model non-Markovian. Indeed, a minimal Markovian representation of the RW is the couple $(X_t, l_t)$ where $X_t$ is the position of the RW at time $t$, and $l_t$ the current length traveled since the beginning of the last jump. Note that while this model is not strictly scale-invariant due to the introduction of a finite speed $1$, we expect our analytical preditions to be valid in the limit of a large number of visited sites.
    \item \textit{TSAW.} This is an example of self-interacting RW \cite{Davis:1990,Barbier:2022,Toth:1995}. In this model, the RW at position $i$ jumps to a neighbouring site $j=i \pm 1$ with probability depending on the number of times $n_j$ it has visited site $j$, 
    \begin{align}
    p(i\to j)=\frac{\exp \left[ -\beta n_j \right]}{\exp \left[ -\beta n_{i-1}\right]+\exp \left[ -\beta n_{i+1} \right]}
     \end{align}
    where $\beta$ is positive. Actually, the TSAW defines a universality class, encompassing all self-interacting RWs which are repelled by the gradient of their occupation times
    \begin{align}
    p(i\to j)\propto w(n_j - n_i)
     \end{align}
    where $w$ a positive, decreasing function \cite{Toth:1995}.
    \item \textit{PSRW.} This is another example of self-interacting RW \cite{Davis:1990,Barbier:2022,Toth:1996}. In this model, the RW at position $i$ jumps to a neighbouring site $j=i \pm 1$ with probability depending on the number of times $n_j$ it has visited site $j$, 
    \begin{align}
    p(i\to j)=\frac{n_j^{-\alpha}}{n_{i-1}^{-\alpha}+ n_{i+1}^{-\alpha} }
     \end{align}
    where $\alpha$ is positive.
    \item \textit{Run-and-tumble particle (RTP).} The run-and-tumble particle is a celebrated model of an active particle, going in straight lines at velocity $v$ and changing the sign of its velocity at random, exponentially-distributed times with rate $\gamma$ \cite{solonActiveBrownian}. 
\end{itemize}

We summarize the properties of the non-Markovian RWs of interest in the Table \ref{tab:recap} (see also \cite{Levernier:2018}).

\begin{table*}[th!]
\begin{tabular}{@{\hspace{0.5cm}}c@{\hspace{0.5cm}}|@{\hspace{0.5cm}}c@{\hspace{0.5cm}}|@{\hspace{0.5cm}}c@{\hspace{0.5cm}}|@{\hspace{0.5cm}}c@{\hspace{0.5cm}}|@{\hspace{0.5cm}}c@{\hspace{0.5cm}}}
     Model & $\dw$  & $\theta$  \\  \hline
     fBm & $1/H$ &   $1-H$ \\
     RAP & 2/3 & 1/4 \\
     L\'evy Walk & $\min(\alpha,2)$ & 1/2        \\
     TSAW  & $3/2$ & $1/3$ \\
     PSRW  & $2$   & $1/4$ \\
     RTP  & $2$   & $1/2$ \\
\end{tabular}
\caption{ Summary of the non-Markovian models considered in this study and of their characteristic parameters.  \label{tab:recap} }
\end{table*}

\section{Data description}
In this section, we describe each of the data sets used in this study:

\begin{itemize}

\item \textit{DNA random walk on HUMBMYH7 and HUMTCRADCV sequences \cite{GenBank}}.

\item \textit{Motion of amoeba intracellular vacuoles ($\text{pixels}=106\text{nm}$)}: Intracellular vacuole trajectories within amoebas, analyzed in a $2$D plane with at least $2048$ frames, exhibit a walk dimension close to $1/\dw \approx 0.67$, as estimated in \cite{Krapf:2019}.

\item \textit{Trajectories of telomeres ($\mu\text{m}$)}: Telomere trajectories within the nucleus of untreated U2OS cells, obtained from Ref. \cite{Stadler:2017}, exhibit a walk dimension of $1/\dw \approx 0.25$. We consider trajectories where the mean-square displacement grows as $t^{0.25 \pm 0.05}$, similar to \cite{Krapf:2019}.

\item \textit{Displacement of primary dermal fibroblast cells ($\mu\text{m}$)}: Across $12$ different age groups, ranging from $2$ to $92$ years old, super diffusive trends are observed \cite{Phillip:2021}. The dataset includes $81$ cell trajectories in 2D with $200$ steps for the youngest patient and $61$ trajectories for the oldest, each with the same length.

\item \textit{Financial indices \cite{Bianchi:2013,Mattera:2021}}: We analyze the logarithm of the Standard and Poor's 500 index since '85 (SPX), the Nikkei 225 (N225) and the Dow-Jones index (DJI) since 1985. Their Hurst index hovers near $0.5$ \cite{Bianchi:2013,Mattera:2021}. 
\end{itemize}

\section{Computing the flip probability in datasets}
For computing the probability $\pi_n$ for real datasets, we separate each time series into subtrajectories separated by a number of time steps precised for each time series. Then, flips in the subtrajectory at different indices $n$ are obtained when the time series exceeds  its current range (interval delimited by its maximum and minimum). The number of subtrajectories is chosen such that we obtain flips for $n$ visited sites such that $\pi_n \leq 10^{-2}$. Below, the details for each dataset in the order presented in the main text.

\begin{enumerate}[label=(\alph*)]
    \item HUMBMYH7 DNA RW: 600 subtrajectories separated by 28 steps.\\
    \item HUMTCRADCV DNA RW: 1500 subtrajectories separated by 48 steps for their starts.\\
    \item Amoeba: 205 $2d$ trajectories of length 2048, 20 subtrajectories separated by 102 steps. \\
    \item Fibroblast cells, Patient 1: 81 cells in 2d, 10 subtrajectories separated by 10 steps by trajectory.\\
    \item Fibroblast cells, Patient 2: 61 cells in 2d with 200 steps, 10 subtrajectories separated by 10 steps by trajectory.\\
    \item Telomere: 50 subtrajectories separated by 40 steps.\\
    \item SPX: 1000 subtrajectories separated by 17 steps.\\
    \item N225: 750 subtrajectories separated by 19 steps.\\
    \item DJI: 500 subtrajectories separated by 16 steps.
\end{enumerate}

\section{Processes belonging to class I}
\subsection{Lévy Walks, the RAP and run-and-tumble particles}
In this section, we show that for Lévy Walks (in the crossing convention), the Random Acceleration Process (RAP) and the run-and-tumble particle (RTP), flips are independent of each other. 
Let \( X(t) \) denote the position of the walker and \( V(t) = X(t+1) - X(t) \) its speed at time \( t \). For the RAP, \( V(t) \) is a simple symmetric random walk; for Lévy Walks (resp. the RTP), \( V(t) = \pm 1 \) switches sign at waiting times drawn from a power-law distribution (resp. exponential distribution). For convenience, we assume \( V = 0 \) immediately before any change of sign.

A key property of Lévy Walks, the RAP and the RTP is that the position process \( X(t) \) loses memory of the past trajectory whenever the speed process \( V(t) \) hits zero.

Assume that the walker has visited \( n \) distinct sites and has eccentricity \( z = M/n \), where \( M \) is the maximum site visited. To show that flips are independent, it suffices to show that flips do not depend on the eccentricity $z$, as $z$ is a physical marker of the past number of flips in the trajectory; for example, $z=0$ or $z=1$ if no flip ever occurred. A flip occurs if:
\begin{itemize}
    \item the speed \( V \) is zero upon first reaching site \( M \), and
    \item the walker exits the visited interval \( [-(n-M), M] \) through the left endpoint \( -(n-M) \), starting with \( V = 0 \) at position \( M \).
\end{itemize}
Indeed, if \( V \neq 0 \) at \( M \), the walker immediately steps to \( M + 1 \), thereby exiting on the right.

Due to the loss of memory when $V=0$, we can factor the flip probability as:
\begin{align}
\label{class-i-indep}
&\mathbb{P}\left(V = 0 \text{ when first reaching } M \mid n \text{ sites visited} \right)\nonumber \\
\times \quad
&\mathbb{P}\left( \text{exit through } -(n-M) \mid X_0 = M, V_0 = 0 \right).
\end{align}

Both terms on the right-hand side of Eq.~\eqref{class-i-indep} are independent of \( M \), by translational invariance:

\begin{itemize}
    \item The second term depends only on the length of the interval, not on its eccentricity. Using translational invariance, we have:
    \begin{align}
        &\mathbb{P}\left( \text{exit through } -(n-M) \mid X_0 = M, V_0 = 0 \right)\nonumber \\
    = \quad &\mathbb{P}\left( \text{exit through } 0 \mid X_0 = n, V_0 = 0 \right).
    \end{align}

    \item The first term is also independent of \( M \). The condition that the process visits \( n \) sites and has maximum \( M \) implies that it started at \( -(n - M) \), reached \( M \) without crossing \( -(n - M + 1) \), and was at zero speed at both endpoints. Using again translational invariance at zero speed, this probability is equal to:
    \begin{align}
    \mathbb{P}\bigg(V = 0 \text{ when first reaching } X = n \mid X_0 = 0, V_0 = 0, \text{ no crossing of } -1 \bigg).
    \end{align}
\end{itemize}

Hence, both terms are independent of \( M \), and the flip probability is independent of the eccentricity $z$. This shows that the RAP, Lévy walks and the RTP belong to class I.

\subsection{The saturating self-interacting RW (SATW)}
Here, we show that the SATW \cite{bremontExactPropagators,Toth:1996,Alessandro:2021} belongs to class (I). This model is defined by the probability transition function from site $i$ to site $j$:
\begin{equation}
    p(i \to j) = \begin{cases}
        \frac{1}{2} \text{ if $j$ has already been visited} \\ \frac{1}{1+\phi} \text{ if $j$ has not been visited.}
    \end{cases}
\end{equation}
Because the transition probabilities are modified only at the edge of the visited interval (and behave like those of a simple RW elsewhere), it is clear that flips are all independent from each other. Indeed, only the size of the visited interval matters, there is no impact of the past trajectory on the flip probability.

\subsection{Explicit computation in the case of the run-and-tumble particle}
We check here that our result $\pi_n \sim \frac{\phi}{n}$ indeed holds for processes that are only asymptotically scale-invariant, focusing on the case of the run-and-tumble particle.
Say that the run-and-tumble particle just visited its $n$th site $x_n$. By translation invariance we can assume $x_n=n$. The probability of no flip occurring at the $(n+1)$st site visited is the probability that, starting with positive speed at site $n$, the particle crosses $n+1$ before it crosses $0$. This reduces to the classical splitting probability, which is known explicitly \cite{gueneauRelatingAbsorbing}:
\begin{equation}
    \mathbb{P}(\text{cross $n+1$ before $0$}|n,+) = \frac{\frac{1}{2} + \frac{\gamma}{v} n}{\frac{1}{2} + \frac{\gamma}{v} (n+1) }\Sim{n \to \infty} 1-\frac{1}{n}.
\end{equation}
We thus recover the flip probability and its prefactor 
\begin{equation}
    \pi_n \Sim{n \to \infty} \frac{\phi}{n}, \quad \phi = \dw \theta = 1.
\end{equation}

\section{Computing the prefactor $A$ for class I}
\subsection{Flip probability knowing that no flip ever occurred}
Before computing the prefactor $A$ for all classes, we need an important auxiliary result. In this section, we show that the conditioned probability $\tilde{\pi}_n$ of a flip occurring at the $(n+1)^{\mathrm{st}}$ site visited—under the condition that no flip has occurred previously—scales as
\begin{equation}
    \label{pintildek0}
    \tilde{\pi}_n \sim \frac{\phi}{n} = \frac{\dw \theta}{n}.
\end{equation}

This asymptotic behavior can be deduced from the classical splitting probability $Q_+(n)$: the probability that the walker, starting at the origin, reaches site $n$ before site $-1$. For the walker to reach $n+1$ before $-1$, it must first reach $n$ before $-1$, and then proceed from $n$ to $n+1$ without flipping, conditioned on the absence of flips throughout its entire trajectory. This yields the recursive relation:
\begin{equation}
    \label{pintildek0-0}
    Q_+(n+1) = Q_+(n)\, (1-\tilde{\pi}_n).
\end{equation}

Using the known asymptotic form $Q_+(n) \sim B n^{-\phi}$ \cite{majumdar} for large $n$, we immediately deduce from Eq.~\eqref{pintildek0-0} that
\[
\tilde{\pi}_n(k) = 1-\frac{Q_+(n+1)}{Q_+(n)} \sim \frac{\phi}{n},
\]
thus recovering the claimed result \eqref{pintildek0}. Note that \eqref{pintildek0} actually holds for all classes.
\par 
Because flips are independent for class I, conditioning on the fact that no flips ever occurred has no effect on the flip probability. Therefore, we have, thanks to \eqref{pintildek0}:
\begin{equation}
    \pi_n = \tilde{\pi}_n \sim \frac{\phi}{n}.
\end{equation}
This immediately yields the prefactor $A= \phi = \dw \theta$ for class I processes.

\section{Computing the prefactor $A$ for class II}
\begin{figure}[h!]
    \centering
    \includegraphics[width=.7\textwidth]{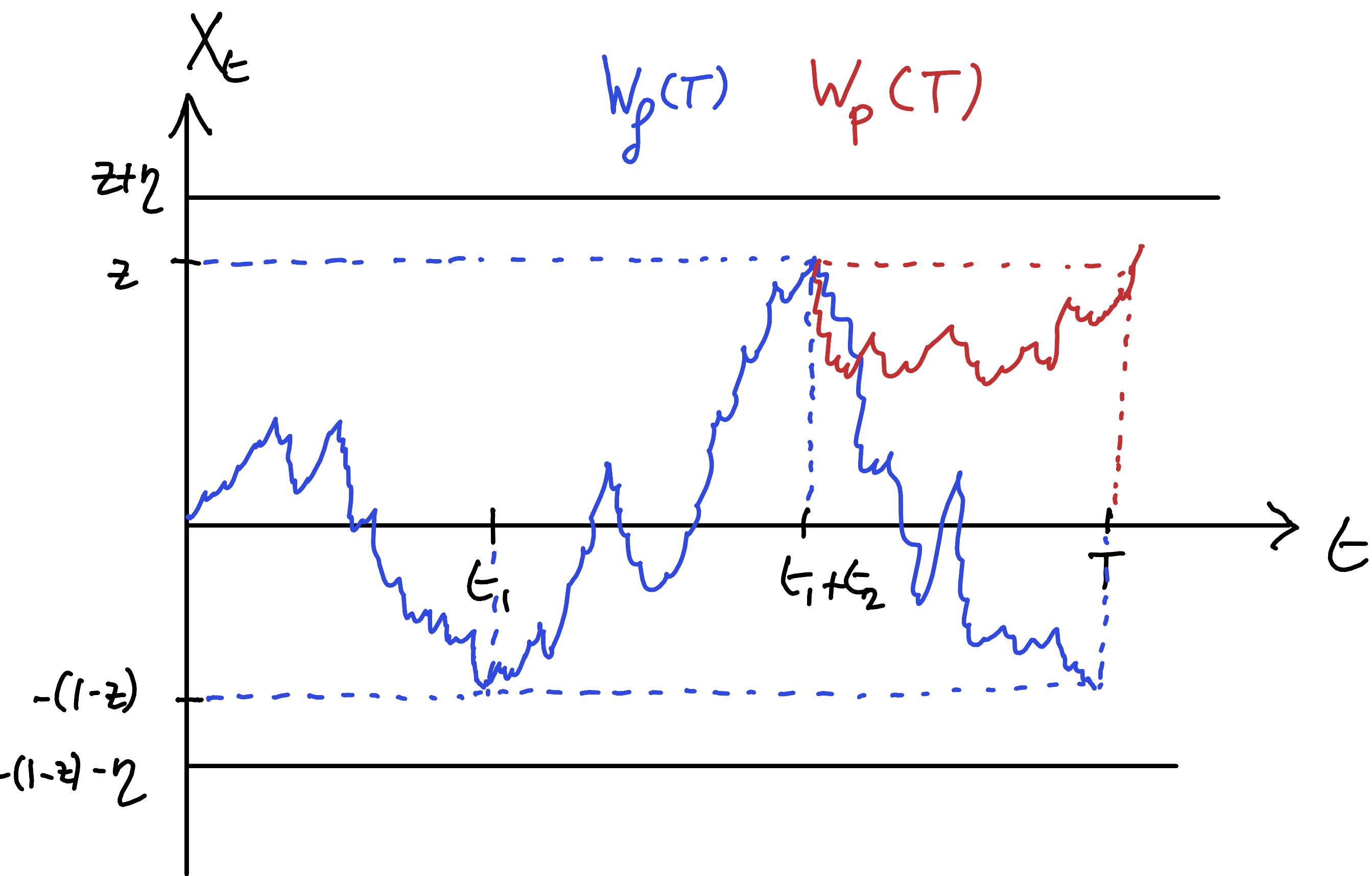}
    \caption{Schematic representation of two fBM paths, blue and red, contributing respectively to the observables $W_f(T)$ and $W_p(T)$.}
    \label{fig:scheme-fbm}
\end{figure}

In this section, we compute analytically to order $\mathcal{O}(\varepsilon^2)$ the prefactor $A$ for the fBM. For this computation, we will use the formalism developed by K.Wiese in \cite{wiese}. This is a diagrammatic, field-theoretical approach based on the explicit expression of the Gaussian action for the fBM at first order in $\varepsilon$. \par \vspace{2ex}
To compute $A$, we introduce two observables: $W_f(T)$ and $W_p(T)$. The first, $W_f(T)$, is defined as the joint probability that the span of the process is $[-(1 - z), \underline{z}]$ and that a flip occurs at time $T$, i.e., the fBM crosses its minimum $-(1 - z)$ at time $T$. The second observable, $W_p(T)$, corresponds to the joint probability that the span is also $[-(1 - z), \underline{z}]$, but the fBM instead crosses its maximum $z$ at time $T$.

To calculate these probabilities, we regularize the problem by introducing absorbing boundaries at $-(1 - z) - \eta$ and $z + \eta$, where the regularization parameter $\eta \to 0$ at the end of the computation. See Fig.~\ref{fig:scheme-fbm} for an illustration. Consider, for example, a path $X_t$ contributing to $W_f(T)$. A crucial observation is that whenever $X_t$ approaches within a distance $\eta$ of either absorbing boundary, its contribution acquires a factor of order $O(\eta)$. To contribute non-trivially to Eq.~\eqref{A-fbm-compute}, such a path must approach the boundaries only three times at $O(\eta)$ distance. For instance, a path contributing to $W_f(T)$ may first cross $-(1 - z)$ at some random time $t_1$, then reach $z$ at $t_1 + t_2$, and finally cross below $-(1 - z) - \eta/2$ at $T = t_1 + t_2 + t_3$. \par

The probability $A_+(z)$ that a flip occurs and that the visited region is $[-(1 - z), \underline{z}]$ (where we underline the last site hit) is then expressed as:
\begin{equation}
    \label{A-fbm-compute}
    A_+(z) = \lim_{\eta \to 0} \frac{W_f}{\eta^3}, W_f = \int_0^\infty W_f(T) \, dT.
\end{equation}
Now, we partition flip events over the eccentricity $z$ and whether the fBM reaches $z$ or $-(1-z)$ first. Thanks to symmetry, we see that the prefactor $A$ writes 
\begin{equation}
    A = \int_0^1 (A_+(z)+A_+(1-z)) \, dz.
\end{equation}
It thus remains to compute the observable $W_f$ at first order in $\varepsilon$. 
At order $0$ in $\varepsilon$, $W_f$ can be obtained simply by the Markov property : at this order, $W_f$ is the product of the probability to hit $-(1-z)$ before $z$ starting from $0$, and of the probability to hit $z$ before $-(1-z)-\eta$ starting from $-(1-z)$. Hence, the probability is simply $z\cdot \eta^3$. We define the leading-order correction to $W_f$ as:
\begin{equation}
    W_f = \eta^3 \left(z + \varepsilon\, \gamma(z) + O(\varepsilon^2)\right).
\end{equation}

Thus, the problem reduces to computing $\gamma(z)$, which corresponds to the first-order correction to the observable $W_f$.

\subsection{The observable $W_f(T)$}
By construction, the observable $W_f(T)$ admits the following path integral representation:
\begin{equation}
    W_f(T) = \int_{X_0 = 0}^{X_T = -(1 - z)} \mathcal{D}X \, \prod_{t = 0}^{T} \left[ \Theta(X_t + 1 - z + \eta)\, \Theta(z + \eta - X_t) \right] \int_{t_1 + t_2 \leq T} \delta(X_{t_1} + 1 - z)\, \delta(X_{t_1 + t_2} - z)\, e^{-S[X]},
\end{equation}
where $S[X]$ is the effective action governing the fBM.

At leading order in $\varepsilon$, this action has been derived in~\cite{wiese} and takes the form:
\begin{equation}
\label{action-exp}
    S = \frac{1}{D} \left(S_0 - \frac{\varepsilon}{2} S_1 + \dots \right), \quad
    D \equiv e^{2\varepsilon(1 + \log \omega)}, \quad
    S_1 \equiv \int_0^{\Lambda} dy \int_0^T dr_1 \int_{r_1}^T dr_2\, \dot{x}(r_1)\, \dot{x}(r_2)\, e^{-y(r_2 - r_1)}.
\end{equation}
Here $\Lambda$ is the large-momentum cutoff, equivalent to the small-time cutoff $\omega$ by $\Lambda = e^{-\gamma}/\omega$. Eventually $\Lambda \to \infty$. An important point is that, from the small-time cutoff, all integrals involving some time $r_1$ as a lower bound actually involve the regularized time $r_1 + \omega$, and hence \eqref{action-exp} involves only the \emph{normal-ordered} velocities $ \dot{x}(r_1) \dot{x}(r_2)$ with $r_2 > r_1$: there is no self-contraction $r_1 = r_2$. \par \vspace{2ex}
From the expressions above, it is clear that we will need to evaluate Brownian velocity correlators of the form $\langle \dot{x}(r_1) \dot{x}(r_2) \rangle$ in the presence of absorbing boundary conditions. However, care must be taken: the relative positions of the times $t_1$, $t_1 + t_2$, $r_1$, and $r_2$ significantly affect the structure of these correlators. For instance, if $r_1 < t_1 < r_2$, then the velocities at $r_1$ and $r_2$ are uncorrelated, because the position $X_{t_1} = z$ is fixed. In this case, the correlator factorizes as
\[
\langle \dot{x}(r_1) \dot{x}(r_2) \rangle = \langle \dot{x}(r_1) \rangle \langle \dot{x}(r_2) \rangle.
\]
Importantly, neither of these averages vanishes: since the endpoints of the path are fixed, the system no longer possesses space-reversal symmetry, and nonzero average velocities can emerge.

To proceed, we introduce the following key functions. First, we define the propagator $Z_t(x_1, x_2)$ as the probability amplitude for a Brownian path to go from $x_1$ at time 0 to $x_2$ at time $t$, in the presence of absorbing boundaries at $-(1 - z) - \eta$ and $z + \eta$. The Laplace transform of this propagator with respect to $t$ is well known \cite{hughes} and will be used extensively.

Second, we define the class of functionals $J_t(u, v; y_1, \dots, y_n)$, which capture the contributions from paths with velocity insertions:
\begin{equation}
    J_t(u,v; y_1,\dots, y_n) = \int_{0 < r_1 < \dots < r_n < t} \prod_{i=1}^n e^{-y_i r_i} \int_{X_0 = u}^{X_t = v} \mathds{1}_{\text{$X$ stays in } [-(1 - z) - \eta,\, z + \eta]} \mathcal{D}X\, \dot{x}(r_1) \dots \dot{x}(r_n)\, e^{-S_0/D}.
\end{equation}

A central tool in the calculations that follow is the Markov property of Brownian paths, which allows us to decompose the path integral over relevant time intervals. This decomposition will be crucial for organizing the contributions to the velocity correlators and computing the first-order corrections to $W_f$.

\subsection{Computing the \texorpdfstring{$J$}{J} functions}

\subsubsection{\texorpdfstring{$J_t(u,v;y)$}{J\_t(u,v;y)}}

By definition, the single-velocity insertion functional reads:
\begin{equation}
    J_t(u,v;y) = \int_0^t e^{-y r} \, dr \int_{X_0 = u}^{X_t = v} \mathds{1}_{\text{$X$ stays in } [-(1 - z) - \eta, z + \eta]} \mathcal{D}X \, \dot{x}(r) \, e^{-S_0/D}.
\end{equation}

We use the known expression for the average velocity of a Brownian path on $[0,t]$, conditioned to start at $u$ and end at $v$:
\begin{equation}
    \langle \dot{x}(r) \rangle = 2D \int_{-\infty}^\infty Z_r(u,x) \, \partial_x Z_{t - r}(x,v) \, dx,
\end{equation}
where $Z_t(x_1,x_2)$ is the propagator in the presence of absorbing boundaries at $-(1 - z) - \eta$ and $z + \eta$.

Taking the Laplace transform in time $t \to s$ yields:
\begin{equation}
    \tilde{J}_s(u,v;y) = 2D \int_{-(1 - z)}^z \tilde{Z}_{s + y}(u,x) \, \partial_x \tilde{Z}_s(x,v) \, dx.
\end{equation}

This function satisfies the symmetry relation:
\begin{equation}
    \tilde{J}_s(u,v;y) = -\tilde{J}_{s + y}(v,u; -y).
\end{equation}

For the case $u < v$, the expression becomes explicitly:
\begin{multline}
    \frac{\tilde{J}_s(u,v;y)}{2D} =
    \frac{\text{csch}(\sqrt{s + y}) \, \sinh\left(\sqrt{s + y}(u - z + 1)\right)}{y}
    \left[\text{csch}(\sqrt{s}) \, \sinh\left(\sqrt{s}(v - z + 1)\right) - \cosh\left(\sqrt{s + y}(v - z)\right)\right] \\
    - \frac{\text{csch}(\sqrt{s}) \, \sinh\left(\sqrt{s}(v - z)\right)}{y}
    \left[\text{csch}(\sqrt{s + y}) \, \sinh\left(\sqrt{s + y}(u - z)\right) + \cosh\left(\sqrt{s}(u - z + 1)\right)\right].
\end{multline}

\subsubsection{\texorpdfstring{$J_t(u,v;y_1, y_2)$}{J\_t(u,v;y1,y2)}}

The two-velocity insertion functional is defined as:
\begin{equation}
    J_t(u,v;y_1,y_2) = \int_0^t dr_1 \int_{r_1}^t dr_2 \, e^{-y_1 r_1 - y_2 r_2} \int_{X_0 = u}^{X_t = v} \mathds{1}_{\text{$X$ stays in } [-(1 - z) - \eta, z + \eta]} \mathcal{D}X \, \dot{x}(r_1)\dot{x}(r_2) \, e^{-S_0/D}.
\end{equation}

We apply the Markov property of Brownian paths to evaluate the velocity correlator for $r_1 < r_2$:
\begin{equation}
    \langle \dot{x}(r_1) \dot{x}(r_2) \rangle = (2D)^2 \int_{-(1 - z) - \eta}^{z + \eta} dx_1 \, dx_2 \, Z_{r_1}(u,x_1) \, \partial_{x_1} Z_{r_2 - r_1}(x_1,x_2) \, \partial_{x_2} Z_{t - r_2}(x_2,v).
\end{equation}

After Laplace transforming (and using $e^{-y_1 r_1 - y_2 r_2} = e^{-(y_1 + y_2) r_1 - y_2 (r_2 - r_1)}$), we obtain:
\begin{equation}
    \tilde{J}_s(u,v;y_1,y_2) = (2D)^2 \int_{-(1 - z) - \eta}^{z + \eta} dx_1 \, dx_2 \, \tilde{Z}_{s + y_1 + y_2}(u,x_1) \, \partial_{x_1} \tilde{Z}_{s + y_2}(x_1,x_2) \, \partial_{x_2} \tilde{Z}_s(x_2,v).
\end{equation}

In practice, we often only require the special case $\tilde{J}_s(u,v;-y,y)$, which simplifies to:
\begin{equation}
    \tilde{J}_s(u,v;-y,y) = (2D)^2 \int_{-(1 - z) - \eta}^{z + \eta} dx_1 \, dx_2 \, \tilde{Z}_s(u,x_1) \, \partial_{x_1} \tilde{Z}_{s + y}(x_1,x_2) \, \partial_{x_2} \tilde{Z}_s(x_2,v).
\end{equation}

When the endpoint $v$ lies near a boundary, for instance at $v = z - \eta$, we use:
\begin{equation}
    \tilde{Z}_s(x_2, z - \eta) = \tilde{Z}_s(x_2, z) - \eta \left. \partial_a \tilde{Z}_s(x_2, a) \right|_{a = z} = -\eta \left. \partial_a \tilde{Z}_s(x_2, a) \right|_{a = z} \equiv \eta \tilde{\mathcal{J}}_s(x_2, z),
\end{equation}
where $\tilde{\mathcal{J}}_s(x_2, z)$ is the Laplace-transformed probability current exiting at the boundary $z$.

Inserting this into the expression above yields:
\begin{equation}
\label{j_y_my}
    \tilde{J}_s(x,z;-y,y) = \eta (2D)^2 \int_{-(1 - z)}^{z} dx_1 \int_{-(1 - z)}^{z} dx_2 \, \tilde{Z}_s(x,x_1) \, \partial_{x_1} \tilde{Z}_{s + y}(x_1,x_2) \, \partial_{x_2} \tilde{\mathcal{J}}_s(x_2, z).
\end{equation}

Equation~\eqref{j_y_my} matches Eq.~(60) of~\cite{wiese}, and is given explicitly in Eq.~(61) of the same reference.

\subsection{Diagrammatic Expansion of \texorpdfstring{$W_f(T)$}{W\_f(T)}}
We identify six distinct diagrammatic contributions to $W_f(T)$, categorized according to the relative positions of the velocity insertion times $r_1 < r_2$ with respect to the key times $t_1$, $t_1 + t_2$, and $T$.

The first three diagrams, denoted $A_1$, $A_2$, and $A_3$, involve correlated velocity insertions:
\begin{itemize}
    \item $A_1$: $0 < r_1 < r_2 < t_1$,
    \item $A_2$: $t_1 < r_1 < r_2 < t_1 + t_2$,
    \item $A_3$: $t_1 + t_2 < r_1 < r_2 < T$.
\end{itemize}
These contributions involve the connected correlator $\langle \dot{x}(r_1) \dot{x}(r_2) \rangle$.

The remaining three diagrams, denoted $B_1$, $B_2$, and $B_3$, involve uncorrelated velocity insertions due to the separation of $r_1$ and $r_2$ into distinct time intervals:
\begin{itemize}
    \item $B_1$: $0 < r_1 < t_1 < r_2 < t_1 + t_2$,
    \item $B_2$: $0 < r_1 < t_1 < t_1 + t_2 < r_2 < T$,
    \item $B_3$: $t_1 < r_1 < t_1 + t_2 < r_2 < T$.
\end{itemize}
In these cases, the velocity correlators factorize: $\langle \dot{x}(r_1) \dot{x}(r_2) \rangle = \langle \dot{x}(r_1) \rangle \langle \dot{x}(r_2) \rangle$.

Fixing the diffusion coefficient to $D = 1$ \footnote{This is the convention chosen by Wiese in his formalism \cite{wiese}, and we keep it here for consistency.}, so that $(2D)^2 = 4$ (since we are not tracking absolute time units and will perform integration over $T$ in the end), the first-order expansion of $W_f(T)$ becomes:
\begin{equation}
    W_f(T) = \eta^3 z - \frac{4\varepsilon}{2} \left(A_1 + A_2 + A_3 + B_1 + B_2 + B_3\right).
\end{equation}

This implies the following expression for the first-order correction $\gamma(z)$:
\begin{equation}
    \gamma(z) = -2\int_0^\infty \frac{A_1 + A_2 + A_3 + B_1 + B_2 + B_3}{\eta^3} \quad dT.
\end{equation}

We now proceed to evaluate each of the six diagrams and extract their leading-order contributions in $\eta$. Since lower-order terms vanish, we focus on terms of order $\eta^3$, which provide the dominant non-zero corrections.

\subsubsection{The \texorpdfstring{$A$}{A} Diagrams}

We now compute the three diagrams $A_1$, $A_2$, and $A_3$, which correspond to contributions from velocity correlators $\langle \dot{x}(r_1)\dot{x}(r_2) \rangle$ when both insertion times lie within the same time interval.

\paragraph{Diagram $A_1$} 
This diagram corresponds to $0 < r_1 < r_2 < t_1$, and reads:
\begin{equation}
    A_1 = \int_{t_1 + t_2 + t_3 = T} \int_0^{\Lambda} dy \, J_{t_1}(0, -(1 - z); -y, y) \, Z_{t_2}(-(1 - z), z) \, Z_{t_3}(z, -(1 - z)) \, dt_1 dt_2 dt_3.
\end{equation}
Taking the Laplace transform in $T \to s$ gives:
\begin{equation}
    \tilde{A}_1(s) = \tilde{Z}_s(-(1 - z), z)^2 \int_0^{\Lambda} \tilde{J}_s(0, -(1 - z); -y, y) \, dy.
\end{equation}
For the time-integrated case ($s = 0$), we use $\tilde{Z}_0(-(1 - z), z) = \eta$ (splitting probability of a Brownian motion starting at distance $\eta$ from the boundary), yielding:
\begin{equation}
    \tilde{A}_1(0) = \eta^2 \int_0^{\Lambda} \tilde{J}_0(0, -(1 - z); -y, y) \, dy.
\end{equation}
This integral is given explicitly in Eq.~(61) of~\cite{wiese}.

\paragraph{Diagram $A_2$} 
For the case $t_1 < r_1 < r_2 < t_1 + t_2$, we write:
\begin{equation}
    A_2 = \int_{t_1 + t_2 + t_3 = T} \int_0^{\Lambda} dy \, Z_{t_1}(0, -(1 - z)) \, J_{t_2}(-(1 - z), z; -y, y) \, Z_{t_3}(z, -(1 - z)) \, dt_1 dt_2 dt_3,
\end{equation}
and in Laplace space:
\begin{equation}
    \tilde{A}_2(s) = \tilde{Z}_s(0, -(1 - z)) \, \tilde{Z}_s(-(1 - z), z) \int_0^{\Lambda} \tilde{J}_s(-(1 - z), z; -y, y) \, dy.
\end{equation}
For $s = 0$, we use $\tilde{Z}_0(0, -(1 - z)) = z$, so:
\begin{equation}
    \tilde{A}_2(0) = z \eta \int_0^{\Lambda} \tilde{J}_0(-(1 - z), z; -y, y) \, dy.
\end{equation}

\paragraph{Diagram $A_3$}
Similarly, for $t_1 + t_2 < r_1 < r_2 < T$, we have:
\begin{equation}
    A_3 = \int_{t_1 + t_2 + t_3 = T} \int_0^{\Lambda} dy \, Z_{t_1}(0, -(1 - z)) \, Z_{t_2}(-(1 - z), z) \, J_{t_3}(z, -(1 - z); -y, y) \, dt_1 dt_2 dt_3,
\end{equation}
leading to:
\begin{equation}
    \tilde{A}_3(s) = \tilde{Z}_s(0, -(1 - z)) \, \tilde{Z}_s(-(1 - z), z) \int_0^{\Lambda} \tilde{J}_s(z, -(1 - z); -y, y) \, dy,
\end{equation}
and at $s = 0$:
\begin{equation}
    \tilde{A}_3(0) = z \eta \int_0^{\Lambda} \tilde{J}_0(z, -(1 - z); -y, y) \, dy.
\end{equation}

\paragraph{Connection to Wiese's Results}
These integrals were evaluated in~\cite{wiese}, where the first-order correction to the outgoing current at the boundary $z$, for a path starting at $x - (1 - z)$, was defined as:
\begin{equation}
    \tilde{\mathcal{A}}(x) = \int_0^{\infty} \tilde{J}_0(x - (1 - z), z; -y, y) \, dy, \quad \text{with } 0 \leq x \leq 1.
\end{equation}
Notably, $\tilde{\mathcal{A}}(0) = \tilde{\mathcal{A}}(1) = 0$. Using this, the $A$ diagrams can be expressed as:
\begin{equation}
    A_1 = -\eta^3 \tilde{\mathcal{A}}(1 - z), \quad A_2 = A_3.
\end{equation}
Explicitly, we have \cite{wiese}
\begin{align}
    \label{atilde}
    \tilde{\mathcal{A}}(x) = (1-2x)\left[12 \log \mathcal{G}-\frac{1}{3} \log 2 \right] - 4 \psi^{(-2)}\left( \tfrac{x+1}{2} \right) 
    + 4 \psi^{(-2)}\left( \tfrac{x}{2} \right) 
    + 4 \psi^{(-2)}\left( 1 - \tfrac{x}{2} \right)
    - 4 \psi^{(-2)}\left( \tfrac{1-x}{2} \right),
\end{align}
where $\mathcal{G}$ is the Glaisher-Kinkelin constant and $\psi$ is the Digamma function \cite{abramowitz1965handbook}.
Here we used the spatial symmetry of the current: $\tilde{\mathcal{A}}(x) = -\tilde{\mathcal{A}}(1 - x)$. In fact, both $A_2$ and $A_3$ diverge in the limit $\Lambda \to \infty$:
\begin{equation}
    \label{a2a3}
    A_2 = A_3 = \eta^3 z \int_0^\Lambda \frac{\left(\coth\left(\sqrt{y}\right) - \text{csch}\left(\sqrt{y}\right)\right)^2}{y} \, dy.
\end{equation}
Indeed, the integrand in \eqref{a2a3} decays as $y^{-1}$ for large $y$, leading to a logarithmic infrared divergence as $\Lambda \to \infty$. However, this divergence poses no problem: as we will show, the diverging  contributions from the $B$ diagrams precisely cancel those from the $A$ diagrams, ensuring a finite result.

\subsubsection{The \texorpdfstring{$B$}{B} Diagrams}

We now compute the diagrams $B_1$, $B_2$, and $B_3$, which correspond to contributions where the velocity insertions $r_1$ and $r_2$ lie in different time intervals. In such cases, the velocity correlator factorizes: $\langle \dot{x}(r_1) \dot{x}(r_2) \rangle = \langle \dot{x}(r_1) \rangle \langle \dot{x}(r_2) \rangle$.

\paragraph{Diagram $B_1$} 
Here, $r_1 \in [0, t_1]$ and $r_2 \in [t_1, t_1 + t_2]$, so we write $r_2 = t_1 + r_2'$, with $r_2' \in [0, t_2]$. The weight in the action contributes an additional exponential factor $e^{-y t_1}$ due to this time separation:
\begin{equation}
    B_1 = \int_{t_1 + t_2 + t_3 = T} \int_0^\Lambda dy \, e^{-y t_1} \, J_{t_1}(0, -(1 - z); -y) \, J_{t_2}(-(1 - z), z; y) \, Z_{t_3}(z, -(1 - z)) \, dt_1 dt_2 dt_3.
\end{equation}
In Laplace space $T \to s$, this becomes:
\begin{equation}
    \tilde{B}_1(s) = \tilde{Z}_s(-(1 - z), z) \int_0^\Lambda \tilde{J}_{s + y}(0, -(1 - z); -y) \, \tilde{J}_s(-(1 - z), z; y) \, dy.
\end{equation}
For $s = 0$, using $\tilde{Z}_0(-(1 - z), z) = \eta$, we obtain:
\begin{equation}
    \tilde{B}_1(0) = \eta \int_0^\Lambda \tilde{J}_y(0, -(1 - z); -y) \, \tilde{J}_0(-(1 - z), z; y) \, dy.
\end{equation}
Applying the symmetry relation $\tilde{J}_y(a,b;-y) = -\tilde{J}_0(b,a;y)$, we rewrite:
\begin{equation}
    \tilde{B}_1(0) = -\eta \int_0^\Lambda \tilde{J}_0(-(1 - z), 0; y) \, \tilde{J}_0(-(1 - z), z; y) \, dy.
\end{equation}
Explicitly, the expression for $\tilde{B}_1(0)$ reads:
\begin{align}
    \label{b1-exp}
    \tilde{B}_1(0) = -\eta^3 \int_0^\Lambda \bigg[ &
    \frac{z \left( \coth(\sqrt{y}) - \text{csch}(\sqrt{y}) \right)^2}{y} \notag \\
    & + \frac{ \left( \coth(\sqrt{y}) - \text{csch}(\sqrt{y}) \right)
    \left( \text{csch}(\sqrt{y}) - \text{csch}(\sqrt{y}) \cosh(\sqrt{y} z) \right)}{y}
    \bigg] dy.
\end{align}

The term proportional to $z$ in \eqref{b1-exp} leads to a divergent integral as $\Lambda \to \infty$. However, this divergence exactly cancels against the contribution from either $A_2$ or $A_3$. 

\paragraph{Diagram $B_2$}
Here, $r_1 \in [0, t_1]$ and $r_2 \in [t_1 + t_2, T]$. Writing $r_2 = t_1 + t_2 + r_2'$, we find two exponential weights: $e^{-y t_1}$ from $r_1$ to $r_2$, and an additional $e^{-y t_2}$ separating the time intervals:
\begin{equation}
    B_2 = \int_{t_1 + t_2 + t_3 = T} \int_0^\Lambda dy \, e^{-y (t_1 + t_2)} \, J_{t_1}(0, -(1 - z); -y) \, Z_{t_2}(-(1 - z), z) \, J_{t_3}(z, -(1 - z); y) \, dt_1 dt_2 dt_3.
\end{equation}
Transforming to Laplace space:
\begin{equation}
    \tilde{B}_2(s) = \int_0^\Lambda \tilde{J}_{s + y}(0, -(1 - z); -y) \, \tilde{Z}_{s + y}(-(1 - z), z) \, \tilde{J}_s(z, -(1 - z); y) \, dy.
\end{equation}
At $s = 0$:
\begin{equation}
    \tilde{B}_2(0) = \int_0^\Lambda \tilde{J}_y(0, -(1 - z); -y) \, \tilde{Z}_y(-(1 - z), z) \, \tilde{J}_0(z, -(1 - z); y) \, dy.
\end{equation}
Applying symmetry to both $\tilde{J}_0$ terms, we obtain:
\begin{equation}
    \tilde{B}_2(0) = \int_0^\Lambda \tilde{J}_0(-(1 - z), 0; y) \, \tilde{Z}_y(-(1 - z), z) \, \tilde{J}_y(-(1 - z), z; -y) \, dy.
\end{equation}
Explicitly, this writes 
\begin{equation}
    \label{b2-exp}
    \tilde{B}_2(0) =\eta ^3  \int_0^\Lambda  \frac{\text{csch}\left(\sqrt{y}\right) \left(z \cosh \left(\sqrt{y}\right)-\cosh \left(\sqrt{y} z\right)-z+1\right)}{\sqrt{y} \left(\cosh \left(\sqrt{y}\right)+1\right)} dy.
\end{equation}
This integral has a finite $\Lambda \to \infty$ limit. 

\paragraph{Diagram $B_3$}
This case involves $r_1 \in [t_1, t_1 + t_2]$ and $r_2 \in [t_1 + t_2, T]$. Writing $r_1 = t_1 + r_1'$, $r_2 = t_1 + t_2 + r_2'$, the weights from the action are $e^{-y r_2 + y r_1} = e^{-y(t_2 + r_2' - r_1')}$, and we find:
\begin{equation}
    B_3 = \int_{t_1 + t_2 + t_3 = T} \int_0^\Lambda dy \, Z_{t_1}(0, -(1 - z)) \, e^{-y t_2} \, J_{t_2}(-(1 - z), z; -y) \, J_{t_3}(z, -(1 - z); y) \, dt_1 dt_2 dt_3.
\end{equation}
In Laplace space:
\begin{equation}
    \tilde{B}_3(s) = \tilde{Z}_s(0, -(1 - z)) \int_0^\Lambda \tilde{J}_{s + y}(-(1 - z), z; -y) \, \tilde{J}_s(z, -(1 - z); y) \, dy.
\end{equation}
Evaluated at $s = 0$, we use $\tilde{Z}_0(0, -(1 - z)) = z$:
\begin{equation}
    \tilde{B}_3(0) = z \eta \int_0^\Lambda \tilde{J}_y(-(1 - z), z; -y) \, \tilde{J}_0(z, -(1 - z); y) \, dy.
\end{equation}
Applying symmetry:
\begin{equation}
    \tilde{B}_3(0) = -z \eta \int_0^\Lambda \tilde{J}_0(z, -(1 - z); y)^2 \, dy.
\end{equation}
Or, in preferred form:
\begin{equation}
    \tilde{B}_3(0) = -z \eta \int_0^\Lambda \tilde{J}_y(-(1 - z), z; -y)^2 \, dy.
\end{equation}
Explicitly, we find:
\begin{equation}
    -z \eta^3 \int_0^\Lambda \frac{\left( \coth(\sqrt{y}) - \text{csch}(\sqrt{y}) \right)^2}{y} \, dy.
\end{equation}
As before, the integral diverges in the infrared limit $\Lambda \to \infty$. However, this divergence is exactly canceled by the corresponding divergence in one of the $A_2$ or $A_3$ diagrams.

\subsection{Summing All Diagram Contributions}

As previously discussed, only the diagrams $A_1, B_1,B_2$ contribute in the $\Lambda \to \infty$ limit. We therefore write:
\begin{align}
    \label{gamma-final}
    \gamma(z) &= -\frac{2}{\eta^3} \left(\tilde{A}_1(0) + \tilde{B}_2(0) + \tilde{B}_1(0)\big|_{\text{finite}} \right) \notag \\
    &= 2\bigg(\tilde{\mathcal{A}}(1-z) -\int_0^\infty \frac{\text{csch}\left(\sqrt{y}\right) 
    \left( z \cosh\left(\sqrt{y}\right) - \cosh\left(\sqrt{y} z\right) - z + 1 \right)}{
    \sqrt{y} \left( \cosh\left(\sqrt{y}\right) + 1 \right)} \, dy \notag \\
    &\hspace{3cm}
    - \int_0^\infty \frac{\left( \coth\left(\sqrt{y}\right) - \text{csch}\left(\sqrt{y}\right) \right) 
    \left( \text{csch}\left(\sqrt{y}\right) - \text{csch}\left(\sqrt{y}\right) \cosh\left(\sqrt{y} z\right) \right)}{y} \, dy.\bigg)
\end{align}

Changing variables via \( y = -\log^2 r \), the first integral in Eq.~\eqref{gamma-final} can be evaluated explicitly. It yields:
\begin{align*}
    &\int_0^\infty \frac{\text{csch}\left(\sqrt{y}\right) 
    \left( z \cosh\left(\sqrt{y}\right) - \cosh\left(\sqrt{y} z\right) - z + 1 \right)}{
    \sqrt{y} \left( \cosh\left(\sqrt{y}\right) + 1 \right)} \, dy \\ &= \frac{1}{2} \bigg( 
        -2 z^2 H_{1-z}
        - 2 z^2 H_{z+1}
        + (2 z^2 - 1) H_{\frac{z+1}{2}} 
        + (2 z^2 - 1) H_{\frac{1}{2} - \frac{z}{2}} 
        + z^2 \log 16 
        - 2z + 4 
        - 2 \log 4 
    \bigg),
\end{align*}
where \( H_z \) denotes the analytic continuation of the harmonic numbers~\cite{abramowitz1965handbook}.

The second integral in Eq.~\eqref{gamma-final} can also be computed using the same substitution. It evaluates to:
\begin{align}
\label{int-b2}
    &\int_0^\infty \frac{\left( \coth\left(\sqrt{y}\right) - \text{csch}\left(\sqrt{y}\right) \right) 
    \left( \text{csch}\left(\sqrt{y}\right) - \text{csch}\left(\sqrt{y}\right) \cosh\left(\sqrt{y} z\right) \right)}{y} \, dy\\  &=-12 \log \mathcal{G} 
    + 4 \psi^{(-2)}\left( \tfrac{z+1}{2} \right) 
    - 4 \psi^{(-2)}\left( \tfrac{z}{2} \right) 
    - 4 \psi^{(-2)}\left( 1 - \tfrac{z}{2} \right)
    + 4 \psi^{(-2)}\left( \tfrac{1}{2} - \tfrac{z}{2} \right) \notag \\
    &\quad - 2z \log \left[ \Gamma \left( 1 - \tfrac{z}{2} \right) \Gamma \left( \tfrac{z+1}{2} \right) \right]
    + 2z \log \left[ \Gamma \left( \tfrac{1}{2} - \tfrac{z}{2} \right) \Gamma \left( \tfrac{z}{2} \right) \right] 
    + \tfrac{1}{3} \log 2 + 2z.
\end{align}
Also, from \cite{wiese}, we have 
\begin{align}
    \label{atilde2}
    \tilde{\mathcal{A}}(1-z) = (2z-1)\left[12 \log \mathcal{G}-\frac{1}{3} \log 2 \right] + 4 \psi^{(-2)}\left( \tfrac{z+1}{2} \right) 
    - 4 \psi^{(-2)}\left( \tfrac{z}{2} \right) 
    - 4 \psi^{(-2)}\left( 1 - \tfrac{z}{2} \right)
    + 4 \psi^{(-2)}\left( \tfrac{1-z}{2} \right).
\end{align}
Summing up, we have 
\begin{align}
    \gamma(z) &= -\frac{2}{3} z \bigg(-72 \log (\mathcal{G})+z \log (64)+3+\log (4)\bigg)+2 z^2 H_{1-z}+2 z^2 H_{z+1}+\left(1-2 z^2\right) H_{\frac{z+1}{2}}+\left(1-2 z^2\right) H_{\frac{1}{2}-\frac{z}{2}}\nonumber \\ &\quad - 4 z \log \left(\frac{\Gamma \left(\frac{1}{2}-\frac{z}{2}\right) \Gamma \left(\frac{z}{2}\right)}{\Gamma \left(1-\frac{z}{2}\right) \Gamma \left(\frac{z+1}{2}\right)}\right)-4+\log (16).
\end{align}
The prefactor $A$ can now be computed:
\begin{equation}
    A = \int_0^1 (A_+(z)+A_+(1-z))\,dz = 1 +2\varepsilon \int_0^1 \gamma(z)\, dz + \mathcal{O}(\varepsilon^2).
\end{equation}
Performing the integral $\int_0^1 \gamma(z)\, dz$ with a mathematical software such as \emph{Mathematica}, we finally obtain the exact first order expansion of $A$ in $\varepsilon$: 
\begin{equation}
    A = 1 - \left(48 \log (\mathcal{G})-4 \gamma -\frac{1}{3} 28 \log (2) \right) \varepsilon + \mathcal{O}(\varepsilon^2),
\end{equation}
where $\gamma$ is Euler's constant. 

\section{Computing the prefactor $A$ for class III}
\subsection{The Brownian Web}
To construct the scaling limit of the TSAW in a mathematically rigorous way, Tóth and Werner introduced a powerful probabilistic object known as the \emph{Brownian web}~\cite{tothTrueSelfrepelling}. Intuitively, the Brownian web is a collection of reflecting and coalescing Brownian paths starting from every point in the upper half-space \(\mathbb{R} \times \mathbb{R}_+\), where \(\mathbb{R}\) represents space and \(\mathbb{R}_+\) represents time. It can be seen as the scaling limit of systems of reflecting and coalescing lattice random walks.

Importantly, the Brownian web gives a complete description of the space-time local time field \( (L_t(x))_{x \in \mathbb{R},\, t > 0} \) associated with a TSAW trajectory. Specifically, the time \( L_x(t) \) that the TSAW has spent at site \( x \) up to time \( t \) is given by the height at position \( x \) of a particular Brownian path in the web. This path the unique one for which the area between it and the real line \(\mathbb{R}\) equals \( t \). Uniqueness is ensured by the coalescing-reflecting structure of the web. For more details on the Brownian Web, we refer the reader to the original and elegant presentation in~\cite{tothTrueSelfrepelling}.
\par 
\vspace{2ex}
For our purposes, the Brownian web provides a concrete and tractable framework to study aging in the TSAW. Its structure—an ensemble of coalescing and reflecting Brownian trajectories—naturally encodes the full history of the walker's past visits. This makes it especially well-suited to capturing the long-range temporal correlations and memory effects that underlie aging. Although initially introduced for mathematical rigor, the Brownian web offers a physically transparent picture of how the TSAW explores space while preserving detailed information about its entire trajectory.

\subsection{Computation of $A_+(z)$ from the Brownian Web}
\begin{figure}[h!]
    \centering
    \includegraphics[width=.7\textwidth]{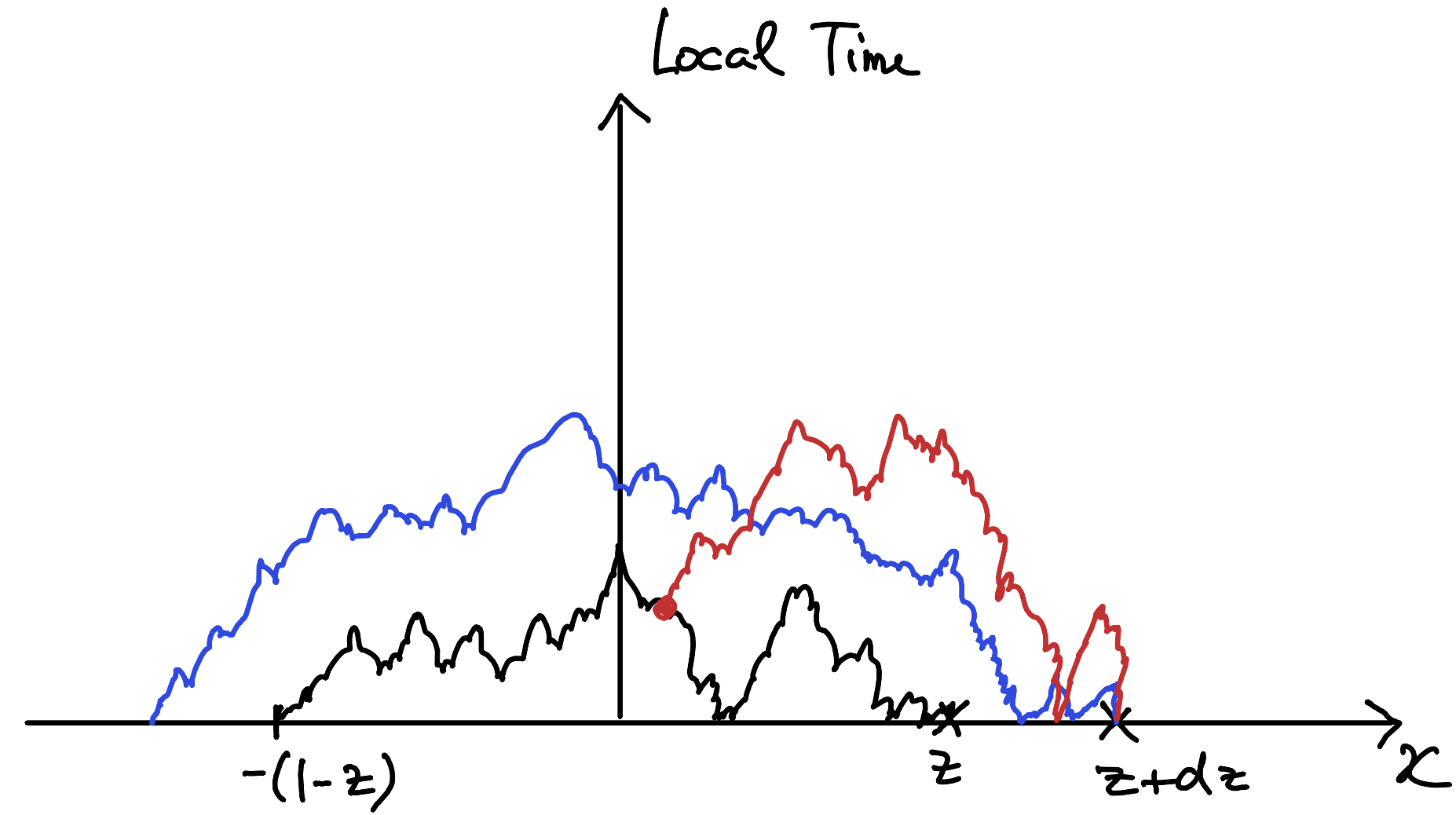}
    \caption{Schematic representation of the computation of the prefactor $A$ for the TSAW using the Brownian web. The black curve shows the initial local time profile when the walker first reaches site \( z \), extending left to \( -(1-z) \), with the initially visited interval rescaled to length 1. Its distribution is given by the Ray--Knight theorem: a reflecting Brownian motion from \( z \) to 0 (right) and a Brownian motion conditioned to hit 0 at \( -(1-z) \), propagating backward from 0 (left).
    The red and blue curves represent possible local time profiles when the walker reaches \( z + dz \). The red path coalesces with the black one before time \( -(1-z) \), indicating that no flip occured when extending the range. On the contrary, the blue path does not coalesce with the initial black profile, meaning the walker flipped—i.e., it reached \( -(1-z)-dz \) before \( z + dz \).
    Crucially, the Brownian web ensures that the red and blue future profiles are Brownian motions starting at \( z + dz \), reflected at 0 within the narrow interval \( [z, z + dz] \), and propagating backward in time until they either coalesce with the black initial profile or are absorbed at the real line. A flip occurs if the future local time profile behaves like the blue one—avoiding coalescence with the initial profile.}
    \label{fig:scheme-tsaw}
\end{figure}
In this section, we present the explicit derivation of the exact prefactor $A$ for class (III), corresponding to non-saturating self-interacting RWs. As we will show, it is sufficient to perform the computation for the TSAW, since all class (III) processes share the same $A$ prefactor. \par 
As in the class (II) case, we will partition flip events over the possible fraction $z\in [0,1]$ of visited sites to the right of the origin. We thus write 
\begin{equation}
    A = \int_0^1 \left(A_+(z)+A_+(1-z) \right) dz,
\end{equation}
where $A_+(z)$ is the joint probability that (i) the TSAW reaches its maximum $z$ when it visits a region of size $1$ and (ii) a flip occurs after the visit of $z$. As explained by the scheme Fig.~\ref{fig:scheme-tsaw}, this probability $A_+(z)$ can be written as
\begin{equation}
    \label{strat-A-tsaw}
    A_+(z) = \lim_{\varepsilon \to 0} \int \mathcal{D}L_1^\varepsilon \, \mathcal{D}L_2 \, \mathds{1}_{\{L_2 \text{ does not coalesce with } L_1^\varepsilon\}}.
\end{equation}
In this expression:
\begin{itemize}
    \item $L_1^\varepsilon(t)$ denotes the initial local time profile at the moment the TSAW reaches \( z \). According to the Brownian Web, this profile is a Brownian motion starting at 0 at time \( t = z \), reflected on the interval \( [0, z] \). When conditioning on the eccentricity being exactly \( z \), it is further conditioned to hit 0 at time \( t = -(1 - z) \). In Eq.~\eqref{strat-A-tsaw}, however, there is no such conditioning. Instead, we integrate over all Brownian paths \( L_1^\varepsilon \) that remain positive on the interval \( [0, -(1 - z)] \) and end at a value \( 0<L_1^\varepsilon(-(1 - z)) \leq \varepsilon \). As in the fBM computation, we remove the regulator at the end by taking the limit \( \varepsilon \to 0 \).
    
    \item $L_2(t)$ denotes the future local time profile at the moment the TSAW reaches \( z + dz \). According to the Brownian web, it is a Brownian motion starting from 0 at time \( t = z + dz \), reflected on the short interval \( [z, z + dz] \), and propagating backward in time until it coalesces with either the initial profile \( L_1(t) \) or the real line—whichever it encounters first. A flip event occurs if it coalesces with the real line, or, equivalently, if it does not coalesce with $L_1^\varepsilon$.
\end{itemize}
At this point, we make a key observation: raising the local time profiles \( L_1^\varepsilon \) and \( L_2 \) in Eq.~\eqref{strat-A-tsaw} to any power \( q > 0 \) leaves the value of $A_+(z)$ unchanged. This holds because:
\begin{enumerate}
    \item All paths start and end at zero, and this remains unchanged when raised to a power \( q \);
    \item Coalescence between \( L_1^\varepsilon \) and \( L_2 \) occurs if and only if the same is true for \( (L_1^\varepsilon)^q \) and \( (L_2)^q \).
\end{enumerate}
The Brownian web constructions for other processes of class (III), such as the PSRW, involve powers of Brownian motions, rather than standard Brownian paths, as shown in~\cite{tothTrueSelfrepelling}. This confirms that the prefactor $A$ is fully universal across all class (III) processes. This argument justifies our focus on the TSAW case.

\subsection{Computation of the Coalescence Probability}
\begin{figure}[h!]
    \centering
    \includegraphics[width=.85\textwidth]{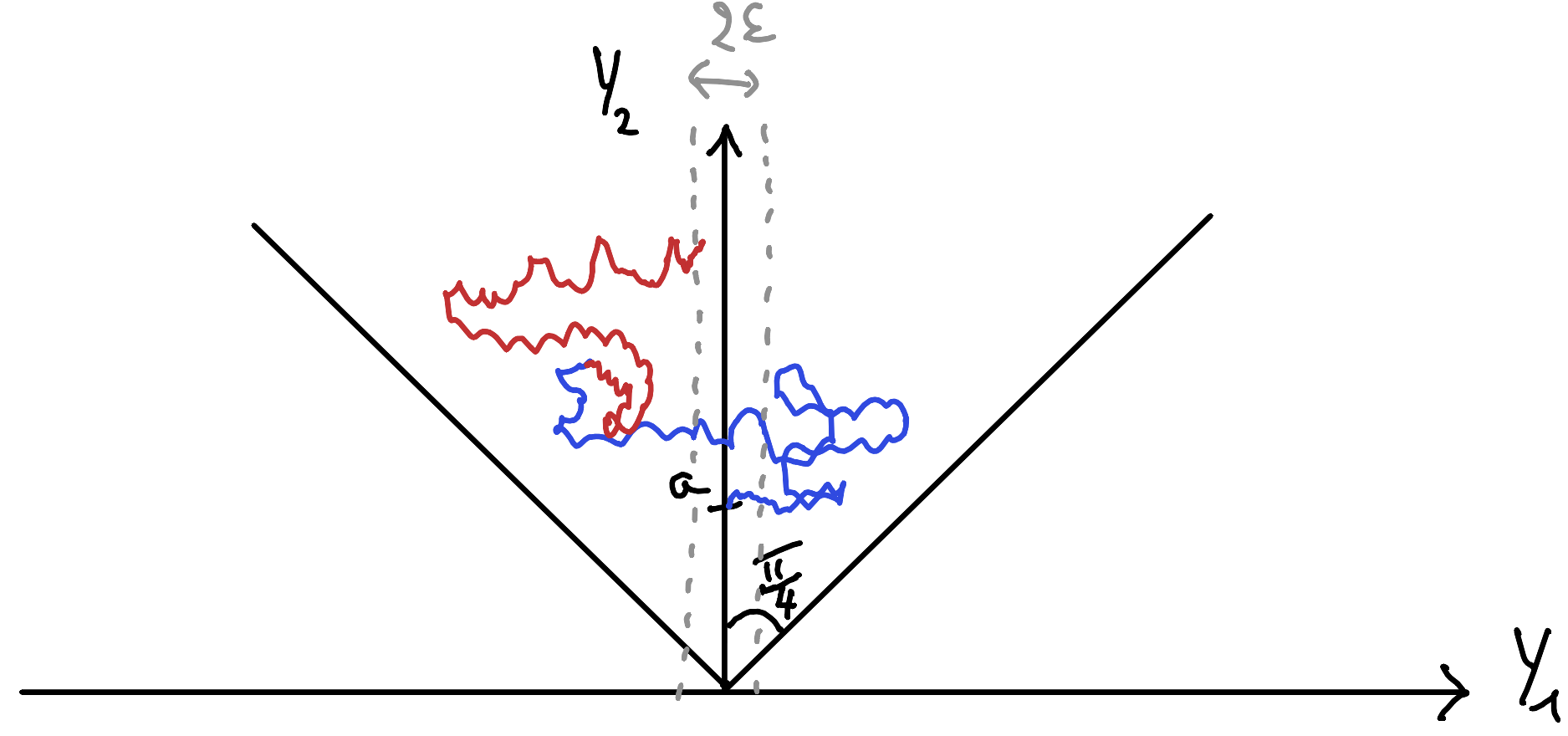}
    \caption{Schematic representation of the mapping to $2d$ Brownian motion. In this figure, the blue path represents the $2d$ Brownian motion $Y(t)$ run from time $t=0$ to time $t=z$. It is started from some point $a$, distributed as $p(a|0,dz)$ where $p(x|y,t)$ is the propagator of the $1d$ reflected Brownian motion. The red path represents $Y(t)$ run from time $t=z$ to time $t=1$. Our aim is now to compute the probability that $Y(t)$, starting from $a$, never hits the wedges of respective angles $\frac{\pi}{2}$ (for $0<t<z$) and $\frac{\pi}{4}$ (for $z<t<1$), and ends in the vertical strip $|Y_1| \leq \varepsilon$ at time $t=1$.
    }
    \label{fig:2dbm-tsaw}
\end{figure}
We now proceed to compute the path integral in Eq.~\eqref{strat-A-tsaw}. To do so, we consider a two-dimensional Brownian motion constructed from the local time profiles:
\begin{equation}
    Y_0(t) = \big( L_1^\varepsilon(t),\,L_2(t)\big).
\end{equation}
Because \( L_2 \) may coalesce with \( L_1^\varepsilon \), the process \( Y_0(t) \) is absorbed when it hits the diagonal \( x = y \).

The time variable \( t \) lies in the interval \( [-dz, 1] \), and the behavior of the Brownian motion \( Y_0(t) \) depends on the specific subinterval:

\begin{itemize}
    \item For \( -dz < t < 0 \), the first coordinate of \( Y_0(t) \) is identically zero, while the second coordinate evolves as a reflected $1d$ Brownian motion started from $0$, since $L_2(t)$ is reflected at $0$ over the interval $[z,z+dz]$.
    
    \item For \( 0 < t < z \), the boundary \( x = 0 \) acts as a reflecting wall, since \( L_1^\varepsilon \) is reflected at zero over this interval.
    
    \item For \( z < t < 1 \), the boundary \( x = 0 \) becomes absorbing, as \( L_1^\varepsilon \) remains strictly positive and is no longer reflected at zero.
\end{itemize}
\par \vspace{2ex}
To simplify the analysis, we remove the reflecting boundary in the first time window by identifying $Y_0$ with its mirror image across the \( x = 0 \) axis. This leads to an equivalent two-dimensional Brownian motion \( Y(t) \) with the following properties:
\begin{itemize}
    \item It is absorbed when it hits either diagonal \( x = y \) or \( x = -y \). Equivalently, $Y(t)$ is a Brownian motion in an absorbing wedge of angle $\frac{\pi}{2}$.
    \item In the second time window \( z < t < 1 \), it is additionally absorbed at \( x = 0 \). Equivalently, $Y(t)$ is a Brownian motion in an absorbing wedge of angle $\frac{\pi}{4}$.
\end{itemize}
This mapping to $2d$ Brownian motion in wedges is illustrated Fig.~\ref{fig:2dbm-tsaw}. Summing up, we showed that we can write 
\begin{equation}
    \label{A-tsaw-propag}
    \begin{aligned}
    A_+(z) 
    &= \lim_{\varepsilon \to 0} \int_0^\infty p(a\,|\,0, dz) \cdot 
        \mathbb{P}^a\left(
            \text{$Y(t)$ ends in the vertical strip $|x| \leq \varepsilon$ at $t=1$ and avoids the wedges}
        \right)
     \, da \\
    &= \lim_{\varepsilon \to 0} 
    \int_0^\infty p(a\,|\,0, dz) 
    \cdot 
        \int_0^\infty db \int_0^\infty dx \int_x^\infty dy \, 
        \Tilde{p}_{\frac{\pi}{2}}\left((x, y) \,|\, (0, a), z\right)
        \Tilde{p}_{\frac{\pi}{4}}\left((\varepsilon, b) \,|\, (x, y), 1 - z\right)
     \, da.
    \end{aligned}
\end{equation}

In Eq.~\eqref{A-tsaw-propag}, we denoted by $p(x\,|\,y, t)$ the propagator of one-dimensional reflected Brownian motion, and by $\tilde{p}_{\theta}((x_2, y_2)\,|\,(x_1, y_1), t)$ the propagator of two-dimensional Brownian motion in an absorbing wedge of angle $\theta$. 

We will now derive the explicit formulae for $2d$ Brownian motion in an absorbing wedge using the image method.

\subsection{The image method}

Let us define the free propagator of two-dimensional Brownian motion as 
\[
G(x, y, t) = \frac{e^{-\frac{x^2 + y^2}{2t}}}{2\pi t}.
\]

Using the image method, the propagator in a $\frac{\pi}{2}$ absorbing wedge is given by
\[
\tilde{p}_{\frac{\pi}{2}}((a,b)\,|\,(x,y),t) = G(a - x, b - y, t) + G(a + x, b + y, t) - \left[ G(a - y, b - x, t) + G(a + y, b + x, t) \right].
\]

For the $\frac{\pi}{4}$ absorbing wedge, we have:
\begin{align*}
\tilde{p}_{\frac{\pi}{4}}((a,b)\,|\,(x,y),t) &= G(a - x, b - y, t) + G(a + x, b + y, t) + G(a - y, b + x, t) + G(a + y, b - x, t) \\
&\quad - \left[ G(a - x, b + y, t) + G(a + x, b - y, t) + G(a - y, b - x, t) + G(a + y, b + x, t) \right].
\end{align*}

From this, we obtain the small-$\varepsilon$ expansion:
\[
\tilde{p}_{\frac{\pi}{4}}((\varepsilon,b)\,|\,(x,y),1-z) \underset{\varepsilon \to 0}{\sim} \varepsilon \cdot \frac{e^{-\frac{b^2 + x^2 + y^2}{2(1-z)}} \left(2x \sinh\left(\frac{b y}{1-z}\right) - 2y \sinh\left(\frac{b x}{1-z}\right)\right)}{\pi (1-z)^2}.
\]

Since the random initial value $a$ satisfies $a \sim dz^2$, it is small. Thus, we can expand the $\tilde{p}_{\frac{\pi}{2}}$ propagator around the point $(0,a\ll \sqrt{z})$:
\[
\tilde{p}_{\frac{\pi}{2}}((x, y)\,|\,(0, a), z) \underset{a^2 \ll z}{\sim} -a^2 \cdot \frac{(x^2 - y^2) e^{-\frac{x^2 + y^2}{2z}}}{2\pi z^3}.
\]

The integrals in Eq.~\eqref{A-tsaw-propag} can now be computed exactly using \textit{Mathematica}. After integration of $A_+(z)$ over the eccentricity $z$, this yields the exact prefactor $A$, valid for all class (III) processes:
\begin{equation}
    A = \frac{4}{\pi^2}.
\end{equation}

\section{Checking the $1/n$ scaling of the flip probability in the alternative definition for $1d$ non-Markovian models}
In this section, we numerically verify that the alternative definition of flips—based on a transition from a newly visited site to the origin, rather than from one extremum to another—also leads to a flip probability that decays as $1/n$. This universal scaling is predicted by the physical argument presented in Eq.~(3) of the main text.

To compute this alternative observable, we record every time $t_n$ at which the size of the explored region increases from $n$ to $n+1$. If, during the interval $[t_n, t_{n+1}]$, the walker crosses the origin $0$, we count this as a flip.

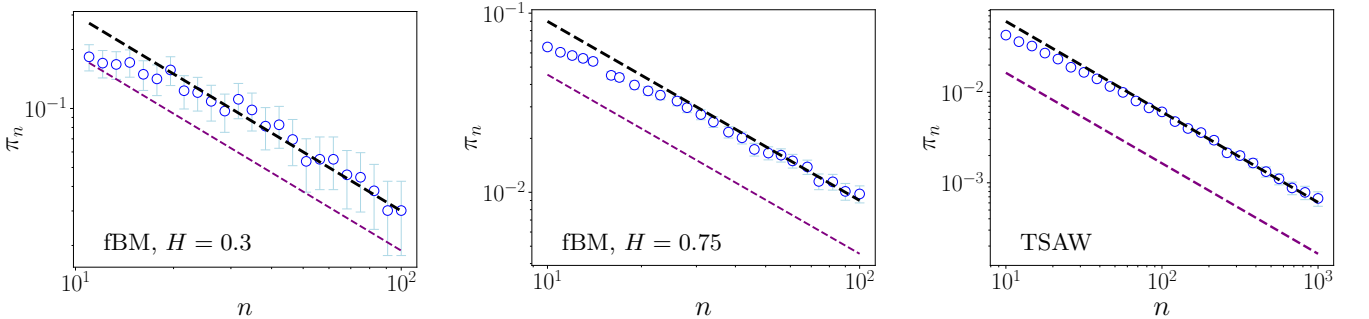
\begin{figure}[h!]
    \centering
    \begin{minipage}[t]{\textwidth}
        \centering
        \begin{subfigure}[t]{0.32\textwidth}
            \centering
            \includegraphics[width=\textwidth]{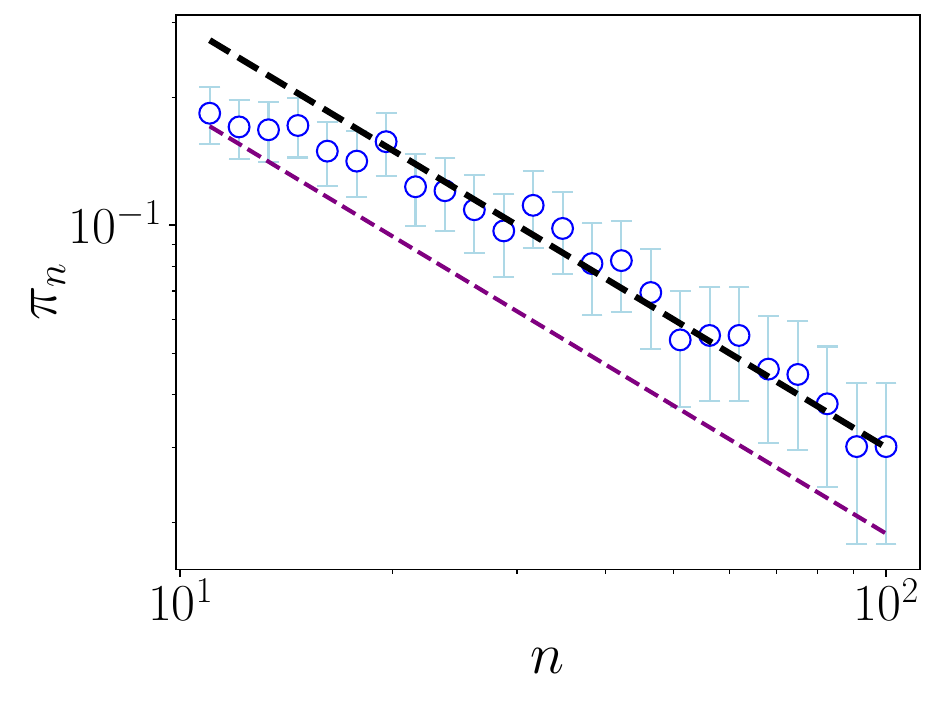}
            \begin{picture}(0,0)
                \put(-38,40){fBM, $H=0.3$}
            \end{picture}
            \label{fig:altdef1}
        \end{subfigure}
        \hfill
        \begin{subfigure}[t]{0.32\textwidth}
            \centering
            \includegraphics[width=\textwidth]{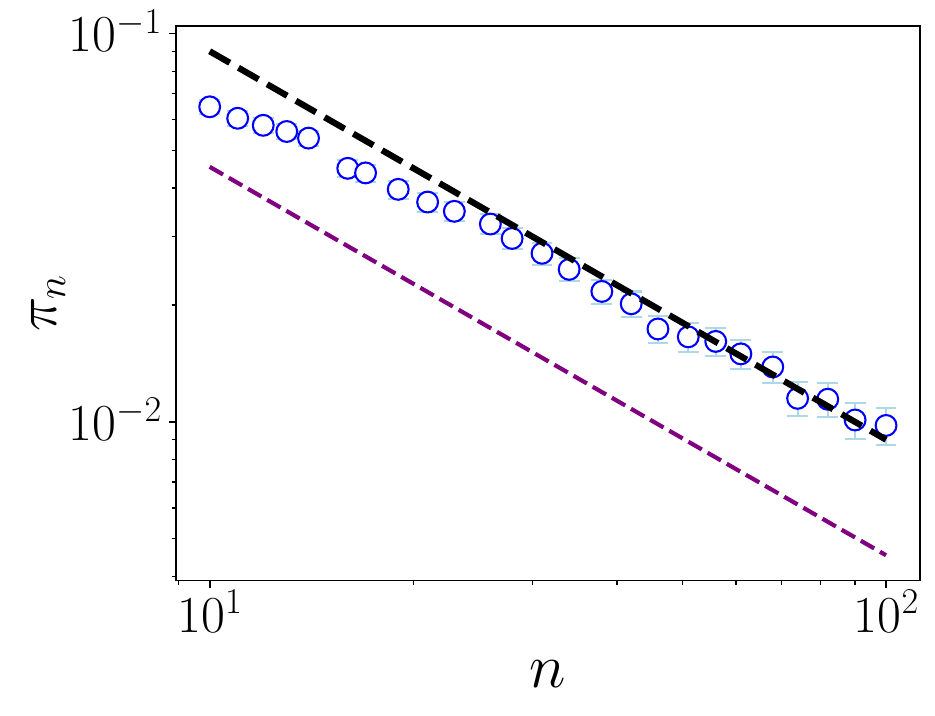}
            \begin{picture}(0,0)
                \put(-38,40){fBM, $H=0.75$}
            \end{picture}
            \label{fig:altdef2}
        \end{subfigure}
        \hfill
        \begin{subfigure}[t]{0.32\textwidth}
            \centering
            \includegraphics[width=\textwidth]{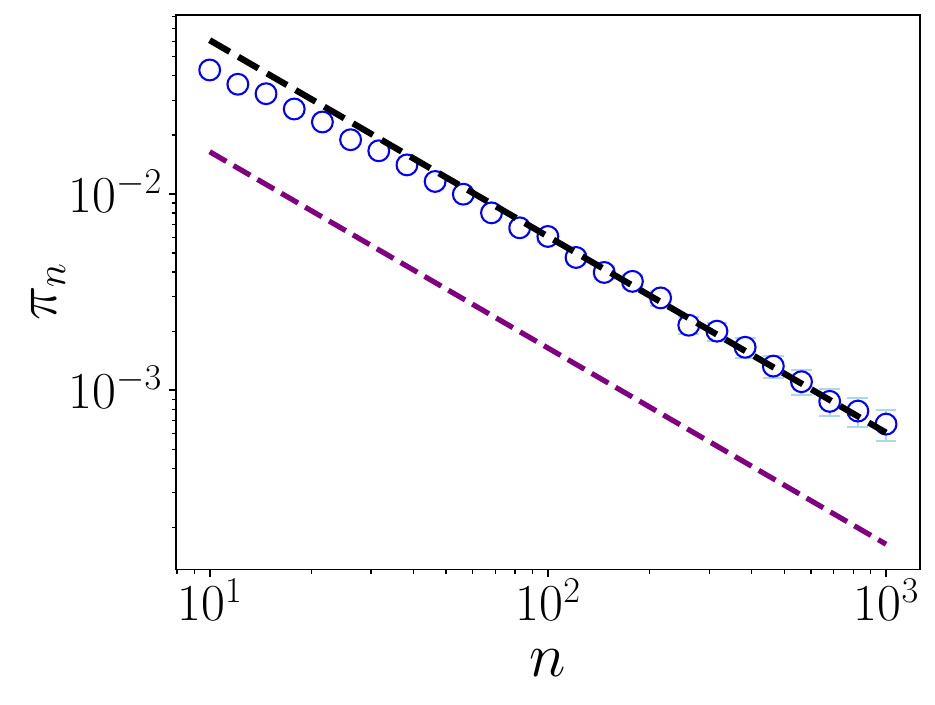}
            \begin{picture}(0,0)
                \put(-38,40){TSAW}
            \end{picture}
            \label{fig:altdef3}
        \end{subfigure}
    \end{minipage}
    \caption{Scaling of the flip probability $\pi_n$ under the alternative definition, for representative non-Markovian models. The black curve corresponds to the numerically measured alternative flip probability, fitted with a $1/n$ scaling. The purple curve shows $A/n$, with $A$ the prefactor obtained from the original flip definition (Eq.~(1), main text). As expected, the alternative definition yields a higher flip probability, since every flip under the original definition (from one extremum to another) necessarily implies a flip under the alternative one (from one extremum to the origin).}
    \label{fig:splitting-alternative}
\end{figure}

\section{Scaling of $\pi_n$ for marginal and transient RWs}
\subsection{Marginal RWs}
In the marginal case, we focus on the $2d$ simple random walk. We evaluate the following probabilities:

\begin{itemize}
    \item[$p_n^1$:] the probability that the $n$th visited site lies on the boundary of the largest completely visited disc $D_n$ of radius $R_n \propto \sqrt{\log n}$ centered at the origin~\cite{demboCoverTimes},
    
    \item[$p_n^0$:] the probability that, starting from a point on the boundary of $D_n$, the walker reaches the origin before exiting $D_n$.
\end{itemize}

We define $p_n = p_n^0 \times p_n^1$. This quantity provides a lower bound for the flip probability $\pi_n$, for the following reason: to observe a flip at the $(n+1)$st visit, the walker must return to the origin from its $n$th visited site $x_n$ without visiting any new site. One possible scenario for this to happen is if $x_n$ lies on the boundary of the fully visited disc $D_n$, so that the path to the origin is already fully explored—maximizing the chance of a flip.

We assume that the boundary of $D_n$ typically contains a non-vanishing fraction of unvisited sites. Therefore, when the walker crosses this boundary, there remains a constant probability of visiting a new site. Thus, the probability of observing a flip is bounded below by
\[
\pi_n \gtrsim p_n^0 \times p_n^1.
\]
This bound captures a specific, feasible mechanism for a flip to occur. While other paths may contribute to $\pi_n$, this construction ensures that the flip probability cannot decay faster than the product $p_n^0 \times p_n^1$.

To estimate $p_n^1$, we integrate the density of position $p(x,t) \propto \frac{e^{-x^2/2t}}{t}$ over the perimeter of the disc $D_n$, which has radius $R_n \propto \sqrt{\log n}$. The number of visited sites scales with time as
\[
n(t) \propto \frac{t}{\log t} \quad \Rightarrow \quad t \propto n \log n.
\]
We thus obtain
\begin{equation}
    p_n^1 \propto \frac{\sqrt{\log n}}{t} \propto \frac{1}{n \sqrt{\log n}}.
\end{equation}

The probability $p_n^0$ is the splitting probability for a $2d$ simple random walk to reach the origin before exiting a disc of radius $R_n = \sqrt{\log n}$, which is well known~\cite{hughes}:
\begin{equation}
    p_n^0 \propto \frac{1}{\log R_n} \propto \frac{1}{\log \log n}.
\end{equation}

Combining these results, we obtain the following lower bound for the flip probability:
\begin{equation}
    \pi_n^{\text{marginal}} \gtrsim \frac{1}{n \sqrt{\log n} \log \log n}.
\end{equation}

This lower bound is close—yet strictly larger—than the simpler bound used in the main text:
\begin{equation}
    \pi_n^{\text{marginal}}  \gtrsim \frac{1}{n \log n}.
\end{equation}

\subsection{Transient RWs}
In the transient case and after $n$ sites have been visited, the radius of the maximally visited ball $B_n$ containing the origin does not scale with $n$ and stays $O(1)$ \cite{demboCoverTimes}. Intuitively, this is because the number of returns to the origin is finite. Hence, when visiting the $n$th site on the boundary of $B_n$, the probability to hit the origin before visiting the $(n+1)$st site cannot scale with $n$, and is again $O(1)$. Therefore, a lower bound for the flip probability $\pi_n$ is the product $p_n^0 \times p_n^1$, where 
\begin{itemize}
    \item[$p_n^1$:] the probability that the $n$th visited site lies on the boundary of the largest completely visited ball $B_n$ of radius $R_n = O(1)$ centered at the origin. $p_n^1$ has the same scaling as the probability $\frac{1}{t^{\dw/\df}} = \frac{1}{t^{1/\mu}}$ to be at the origin at the time $t_n$ at which the $n$th site is discovered. It is well-known \cite{hughes} that $t_n \propto n$ for transient RWs, and hence $p_n^1 \propto \frac{1}{n^{1/\mu}}$.
    
    \item[$p_n^0$:] the probability that, starting from a point on the boundary of $B_n$, the walker reaches the origin before exiting $B_n$. As implied earlier, $p_n^0 = \mathcal{O}(1)$.
\end{itemize}
We therefore obtain the lower bound:
\begin{equation}
    \pi_n^{\text{transient}}  \gtrsim p_n^0 \times p_n^1 \propto \frac{1}{n^{1/\mu}}.
\end{equation}

\end{document}